\documentclass[aps,prd,a4paper,twocolumn]{revtex4}

\usepackage{graphicx}% Include figure files
\usepackage{bm}% bold math
\usepackage{url}
\usepackage{amsfonts}
\usepackage{amsmath}

\newcommand{\muas}[0]{\hbox{\rm $\mu$as}}

\newcommand{\ve}[1]{\mbox{\boldmath$#1$}}

\begin{document}

\title{Light propagation in the gravitational field of one arbitrarily moving pointlike body \\ in the 2PN approximation}   

\author{Sven \surname{Zschocke}}
\affiliation{
Institute of Planetary Geodesy - Lohrmann Observatory, Dresden Technical University,\\
Helmholtzstrasse 10, \\
 D-01069 Dresden, Germany\\
}

\begin{abstract}

An analytical solution for the light trajectory in the near-zone of the gravitational field of one pointlike body in arbitrary slow-motion in the 
post-post-Newtonian approximation is presented in harmonic gauge. Expressions for total light deflection and time delay are given.  
The presented solution is a further step toward high-precision astrometry aiming at nano-arcsecond level of accuracy.  

\end{abstract}

%------------------------------------------------------------

% \date{\today}

\maketitle

%------------------------------------------------------------

%%%%%%%%%%%%%%%%%%%%%%%%%%%%%%%%%%%
\section{Introduction}\label{Section1}  
%%%%%%%%%%%%%%%%%%%%%%%%%%%%%%%%%%%

In order to determine the positions and motions of astronomical objects on the sky, astrometry uses light signals (photons) which are emitted by the 
celestial objects. These light rays propagate from the celestial light source through the gravitational  
field of the Solar System and do finally arrive at the observer. Therefore, the precise determination of the trajectories of light signals  
through the warped space-time of Solar System is a fundamental assignment of a task in relativistic astrometry. According to the theory of general  
relativity \cite{Einstein1,Einstein2} light rays propagate along null-geodesics governed by the geodesic equation,  
\begin{eqnarray}
\frac{d^2 x^{\alpha}\left(\lambda\right)}{d \lambda^2}
+ \Gamma^{\alpha}_{\mu\nu}\,\frac{d x^{\mu}\left(\lambda\right)}{d \lambda}\,
\frac{d x^{\nu}\left(\lambda\right)}{d \lambda} &=& 0\,,
\label{Geodetic_Equation1}
\\
\nonumber\\
g_{\alpha\beta}\,\frac{d x^{\alpha}\left(\lambda\right)}{d \lambda}\,\frac{d x^{\beta}\left(\lambda\right)}{d \lambda} &=& 0\,,
\label{Isotropic_Condition1}
\end{eqnarray}

\noindent
where (\ref{Geodetic_Equation1}) represents the geodesic equation and the constraint (\ref{Isotropic_Condition1}) must be imposed for null-geodesics 
which states that the tangent four-vector along light rays is isotropic.  
In (\ref{Geodetic_Equation1}) and (\ref{Isotropic_Condition1}) the four-coordinates of a light-signal $x^{\alpha}\left(\lambda\right)$ depend on  
affine parameter $\lambda$, and the Christoffel symbols in (\ref{Geodetic_Equation1}) are related to the metric $g_{\alpha\beta}$ of curved space-time,   
\begin{eqnarray}
\Gamma^{\alpha}_{\mu\nu} &=& \frac{1}{2}\,g^{\alpha\beta}
\left(\frac{\partial g_{\beta\mu}}{\partial x^{\nu}}
+ \frac{\partial g_{\beta\nu}}{\partial x^{\mu}}
- \frac{\partial g_{\mu\nu}}{\partial x^{\beta}}\right),
\label{Christoffel_Symbols}
\end{eqnarray}

\noindent
with metric signature $\left(-,+,+,+\right)$.  
The geodesic equation (\ref{Geodetic_Equation1}) and the isotropic condition (\ref{Isotropic_Condition1}) are valid in any reference system. With the aid of  
the zeroth component of (\ref{Geodetic_Equation1}), the geodesic equation and the isotropic condition can be expressed in terms of coordinate time $t$ rather 
than the affine parameter $\lambda$ as follows \cite{MTW,Brumberg1991,Book_Clifford_Will},  
\begin{eqnarray}
\frac{d^2 x^{i}\left(t\right)}{c^2 dt^2}
+ \Gamma^{i}_{\mu\nu} \frac{d x^{\mu}\left(t\right)}{c dt}
\frac{d x^{\nu}\left(t\right)}{c dt}
&=& \Gamma^{0}_{\mu\nu} \frac{d x^{\mu}\left(t\right)}{c dt} \frac{d x^{\nu}\left(t\right)}{c dt} \frac{d x^{i}\left(t\right)}{c dt},
\nonumber\\
\label{Geodetic_Equation2}
\\
g_{\alpha\beta}\,\frac{d x^{\alpha}\left(t\right)}{c dt}\,\frac{d x^{\beta}\left(t\right)}{c dt} &=& 0\,,   
\label{Isotropic_Condition2}
\end{eqnarray}

\noindent
while the zeroth component in (\ref{Geodetic_Equation2}) vanishes identically. The equations in (\ref{Geodetic_Equation2}) and (\ref{Isotropic_Condition2})  
are more appropriate in order to integrate the geodesic equation and also in view of the fact that real astrometric measurements do by all means imply the use 
of concrete reference systems.  
In line with the resolutions of International Astronomical Union (IAU) \cite{IAU_Resolution1}, the Barycentric Celestial Reference System 
(BCRS) is adopted, which is the standard global chart in modern-day astrometry. The origin of the spatial axes of the BCRS is located at the barycenter 
of the Solar system, the harmonic coordinates of the BCRS are denoted by $\left(ct,x^i\right)$ where $t$ is the BCRS coordinate time and $x^i$ are the  
three-dimensional coordinates referred to the spatial axes of the BCRS, and obey the harmonic gauge condition (de Donder gauge):   
\begin{eqnarray} 
\frac{\partial \sqrt{-g}\,g^{\alpha\beta}}{\partial x^{\alpha}} &=& 0\,, 
\label{Harmonic_Gauge}
\end{eqnarray}

\noindent 
where $g = {\rm det}\left(g_{\mu \nu}\right)$ is the determinant of metric tensor.  

For a unique solution of the geodesic equation (\ref{Geodetic_Equation2}) mixed initial-boundary conditions must be imposed  
\cite{Brumberg1987,Brumberg1991,KlionerKopeikin1992,Klioner_Zschocke,Kopeikin1997,KopeikinSchaefer1999_Gwinn_Eubanks,Zschocke_1PN,Zschocke_15PN}:
\begin{eqnarray}
\ve{x}_0 &=& \ve{x}\left(t_0\right),
\label{Initial_Boundary_Condition_1}
\\
\nonumber\\
\ve{\sigma} &=& \lim_{t \rightarrow - \infty}\, \frac{\dot{\ve{x}}\left(t\right)}{c}\,, 
\label{Initial_Boundary_Condition_2}
\end{eqnarray}

\noindent
where the dot in (\ref{Initial_Boundary_Condition_2}) denotes total derivative with respect to coordinate time. 
The first condition (\ref{Initial_Boundary_Condition_1}) defines the spatial coordinates of the photon at the moment
$t_0$ of emission of light. The second condition (\ref{Initial_Boundary_Condition_2}) defines the unit-direction
of the light ray at past null infinity, that means the
unit-tangent vector along the light path in the infinite past hence at infinite spatial distance from the origin of the global coordinate system.
Then, the exact solution of (\ref{Geodetic_Equation2}) for the  
trajectory of the light ray, propagating from the light source through the Solar System towards the observer, can formally be written as follows,
\begin{eqnarray}
\ve{x}\left(t\right) &=& \ve{x}_0 + c \left(t-t_0\right) \ve{\sigma} + \Delta \ve{x}\,,
\label{Introduction_5}
\end{eqnarray}

\noindent
where the term $\Delta \ve{x}$ denotes gravitational corrections to the unperturbed light ray.

In case of weak gravitational fields it is useful to decompose the metric tensor as follows,   
\begin{eqnarray}
g_{\alpha \beta}\left(t,\ve{x}\right) &=& \eta_{\alpha \beta} + h_{\alpha \beta}\left(t,\ve{x}\right), 
\label{metric_perturbation}
\end{eqnarray}

\noindent
where $\eta_{\alpha\beta} = \eta^{\alpha\beta} = {\rm diag}\left(-1,+1,+1,+1\right)$ is the metric of Minkowskian space and for any components  
of the metric perturbations $\left|h_{\alpha\beta}\right| \ll 1$. Because the gravitational fields are weak in the Solar system,  
the orbital motions of the Solar system bodies are slow (virial theorem), $m_A/P_A \ll 1$ and $v_A/c \ll 1$ (notations are given in the Appendix \ref{Appendix_Notation}),  
hence an expansion of the metric in terms of inverse powers of the speed of light can be applied, called post-Newtonian expansion   
or weak-field slow-motion approximation \cite{Brumberg1991,IAU_Resolution1,KlionerKopeikin1992,Blanchet_Faye_Ponsot,DSX1,DSX2,Kopeikin_Efroimsky_Kaplan}, 
which for the covariant and contravariant components reads  
\begin{eqnarray}
g_{\alpha \beta} &=& \eta_{\alpha \beta} + h^{(2)}_{\alpha\beta} + h^{(3)}_{\alpha\beta} + h^{(4)}_{\alpha\beta} + {\cal O} \left(c^{-5}\right),
\label{post_Newtonian_metric_B}
\\
\nonumber\\
g^{\alpha \beta} &=& \eta^{\alpha \beta} - h_{(2)}^{\alpha\beta} - h_{(3)}^{\alpha\beta} - h_{(4)}^{\alpha\beta} + {\cal O}\left(c^{-5}\right),
\label{post_Newtonian_metric_C}
\end{eqnarray}

\noindent
where $h^{(n)}_{\alpha\beta} = {\cal O} \left(c^{-n}\right)$ with $n=2, 3, 4$; e.g. Eqs.~(4.17) - (4.19) in \cite{Kopeikin_Efroimsky_Kaplan}. 
Notice that the post-Newtonian expansion in (\ref{post_Newtonian_metric_B}) 
describes the metric in the near-zone of the Solar System defined by $\left|\ve{x}\right| < \lambda_{\rm gr}$ where $\lambda_{\rm gr}$ is a characteristic  
wavelength of gravitational radiation emitted by the Solar System. It should be mentioned that, according to the famous theorem in \cite{Blanchet_Damour1},  
the post-Newtonian expansion of the metric tensor is, in fact, non-analytic because it contains logarithmic terms. However, in the near-zone the post-Newtonian  
expansion in inverse powers of the speed of light is valid up to 4PN approximation, that means logarithmic terms in metric coefficients emerge 
at the order of ${\cal O}\left(c^{-8}\right)$ \cite{Book_PN}. 
 
The post-Newtonian expansion of the metric in (\ref{post_Newtonian_metric_B}) inherits a corresponding post-Newtonian expansion of the 
light trajectory (\ref{Introduction_5}), which up to terms of the order ${\cal O}\left(c^{-5}\right)$ reads  
\begin{eqnarray}
\ve{x}\left(t\right) &=& \ve{x}_0 + c \left(t - t_0\right) \ve{\sigma} + \Delta \ve{x}_{\rm 1PN} + \Delta \ve{x}_{\rm 1.5PN}
+ \Delta \ve{x}_{\rm 2PN}\,,
\nonumber\\ 
\label{Introduction_6}
\end{eqnarray}
 
\noindent
where the label ${\rm 1PN}$, ${\rm 1.5PN}$, and ${\rm 2PN}$ refer to terms of the order ${\cal O}\left(c^{-2}\right)$, ${\cal O}\left(c^{-3}\right)$,
and ${\cal O}\left(c^{-4}\right)$, respectively. 

The expressions for $\Delta \ve{x}_{\rm 1PN}$ and $\Delta \ve{x}_{\rm 1.5PN}$ for a light trajectory in  
the field of $N$ arbitrarily moving bodies of finite size have recently been determined in \cite{Zschocke_1PN,Zschocke_15PN}. In these investigations each   
individual body $A=1,2,...,N$ is allowed to move along its own arbitrary worldline $\ve{x}_A\left(t\right)$ and the global  
metric of the Solar System has been described in terms of the full set of time-dependent intrinsic mass-multipoles $M^L_A\left(t\right)$ and 
full set of time-dependent intrinsic spin-multipoles $S^L_A\left(t\right)$, 
allowing for arbitrary shape, inner structure and rotational motion of the massive bodies of the Solar System. About the magnitude of these terms 
in time-delay and light deflection we refer to Table II and Table III in \cite{Zschocke_15PN}.   

However, rapidly growing accuracy in astrometric measurements demands to account for post-post-Newtonian terms $\Delta \ve{x}_{\rm 2PN}$ as well.  
In particular, it is well-known that present-day precision in astrometry has reached a level of a few micro-arcseconds ($\muas$)  
in angular observations of stars \cite{GAIA1,GAIA2} and a level of a few nano-seconds (${\rm ns}$) in measurements of time delay \cite{Cassini}. Such extremely  
high-precision astrometry necessitates to account for 2PN effects in the theory of light propagation \cite{KlionerKopeikin1992,Klioner2003a}.  
On the other side, results about the post-post-Newtonian terms $\Delta \ve{x}_{\rm 2PN}$ in (\ref{Introduction_6}) are fairly rare.  
So far, 2PN effects in light propagation have mainly been determined for
the case of mass monopoles at rest \cite{EpsteinShapiro,FischbachFreeman,RichterMatzner1,RichterMatzner2,RichterMatzner3,Cowling}, that means 
where the position of the mass monopole remains constant: $\ve{x}_A = {\rm const}$.  
In this respect, an important progress in calculating post-post-Newtonian effects on light propagation  
in the monopole field has been achieved in \cite{Brumberg1987,Brumberg1991} where
an explicit 2PN solution for light trajectories in the Schwarzschild field as function of coordinate time has been found and later been
confirmed within several progressing investigations \cite{KlionerKopeikin1992,Bruegmann2005,Klioner_Zschocke,Deng_Xie,Deng_2015,Zschocke_15PN}. 
Also alternative approaches for the calculation of directions of light rays and their propagation time
in 2PN approximation have been developed, which avoid the peculiarities of solving the null geodesic equations,
based on the eikonal concept \cite{Ashby_Bertotti}, on the Synge's world function \cite{Poncin_Lafitte_Teyssandier_2008}
or on the Time Transfer Function formalism \cite{Teyssandier,Hees_Bertone_Poncin_Lafitte_2014b}.

An ambitious goal in astrometric measurements in near future is to aim at sub-micro-arcsecond (sub-\muas) level in angular determination and
sub-nano-second (sub-ns) level in time delay measurements. For instance, several space-based astrometry missions are under discussion 
which have been proposed to the European Space Agency (ESA) which aim at precisions on sub-nano-arcsecond (sub-nas) level in angular determination 
of celestial objects \cite{NEAT1,NEAT2,Astrod1,Astrod2,Lator1,Lator2,Odyssey,Sagas}. Such extremely high-precision astrometry needs to account for 
the impact of the motion of massive bodies on the light propagation in 2PN approximation.  
The problem, however, of light propagation in the field of moving monopoles  
in 2PN approximation has not been considered yet, aside from the investigation in \cite{Bruegmann2005} which was not intended  
for the problem of light propagation in the Solar System. For this reason, we will consider the problem of light propagation   
through the gravitational field of one pointlike body in slow but otherwise arbitrary motion in the 2PN approximation.  
The article is organized as follows: In Section \ref{Section2} the geodesic equation in 2PN approximation is presented, 
in Section \ref{Section3} the metric of one massive pointlike body in arbitrary motion in 2PN approximation is given,  
in Section \ref{Section4} and Section \ref{Section5} the first and second integration of geodesic equation is represented. The observable 
effects of total light deflection and time delay are given in Section \ref{Observables}.  
A summary and outlook can be found in Section \ref{Section6}. The notation in use is given in the Appendix \ref{Appendix_Notation}.  
 
%%%%%%%%%%%%%%%%%%%%%%%%%%%%%%%%%%%
\section{Geodesic equation in 2PN approximation}\label{Section2}
%%%%%%%%%%%%%%%%%%%%%%%%%%%%%%%%%%%
 
The Solar System is composed of $N$ massive bodies of finite size which move according their mutual gravitational interaction.
In our investigation we will consider one of these massive bodies and approximate the body as pointlike object with Newtonian rest mass $M_A$.
By inserting the metric (\ref{post_Newtonian_metric_B}) into (\ref{Geodetic_Equation2}) we obtain the geodesic equation in 2PN approximation,  
which in terms of global coordinate time reads \cite{Brumberg1991,KopeikinSchaefer1999_Gwinn_Eubanks,Bruegmann2005}  
\begin{widetext}
\begin{eqnarray}
\frac{\ddot{x}^i \left(t\right)}{c^2} &=& + \frac{1}{2}\,h_{00,i}^{(2)}
- h_{00,j}^{(2)} \frac{\dot{x}^i\left(t\right)}{c}\frac{\dot{x}^j\left(t\right)}{c}
- h_{ij,k}^{(2)}\,\frac{\dot{x}^j\left(t\right)}{c}\frac{\dot{x}^k\left(t\right)}{c}
+ \frac{1}{2}\,h_{jk,i}^{(2)}\,\frac{\dot{x}^j\left(t\right)}{c}\frac{\dot{x}^k\left(t\right)}{c}
- h_{ij,0}^{(2)} \frac{\dot{x}^j\left(t\right)}{c}
\nonumber\\
\nonumber\\
&& \hspace{-1.2cm} + \frac{1}{2}\,h_{jk,0}^{(2)} \frac{\dot{x}^i\left(t\right)}{c}
\frac{\dot{x}^j\left(t\right)}{c}\frac{\dot{x}^k\left(t\right)}{c}
- \frac{1}{2}\,h_{00,0}^{(2)}\,\frac{\dot{x}^i\left(t\right)}{c}
- h_{0i,j}^{(3)} \frac{\dot{x}^j\left(t\right)}{c}
+ h_{0j,i}^{(3)} \frac{\dot{x}^j\left(t\right)}{c}
- h_{0j,k}^{(3)}\frac{\dot{x}^i\left(t\right)}{c}\frac{\dot{x}^j\left(t\right)}{c}\frac{\dot{x}^k\left(t\right)}{c}
\nonumber\\
\nonumber\\
&& \hspace{-1.2cm} - h_{0i,0}^{(3)} - \frac{1}{2}\,h_{ij}^{(2)}\,h_{00,j}^{(2)}
- h_{00}^{(2)}\,h_{00,j}^{(2)}\,\frac{\dot{x}^i\left(t\right)}{c}\,\frac{\dot{x}^j\left(t\right)}{c}
+ h_{is}^{(2)}\,h_{sj,k}^{(2)}\,\frac{\dot{x}^j\left(t\right)}{c}\,\frac{\dot{x}^k\left(t\right)}{c}
- \frac{1}{2}\,h_{is}^{(2)}\,h_{jk,s}^{(2)}\,\frac{\dot{x}^j\left(t\right)}{c}\,\frac{\dot{x}^k\left(t\right)}{c}
\nonumber\\
\nonumber\\
&& \hspace{-1.2cm} + \frac{1}{2}\,h_{00,i}^{(4)}\,
- h_{00,j}^{(4)}\,\frac{\dot{x}^i\left(t\right)}{c}\,\frac{\dot{x}^j\left(t\right)}{c}
- h_{ij,k}^{(4)}\,\frac{\dot{x}^j\left(t\right)}{c}\,\frac{\dot{x}^k\left(t\right)}{c}
+ \frac{1}{2}\,h_{jk,i}^{(4)}\,\frac{\dot{x}^j\left(t\right)}{c}\,\frac{\dot{x}^k\left(t\right)}{c} 
\nonumber\\
\nonumber\\
&& \hspace{-1.2cm} + h_{0j,i}^{(4)}\,\frac{\dot{x}^j\left(t\right)}{c} - h_{0i,j}^{(4)}\,\frac{\dot{x}^j\left(t\right)}{c}
- h_{0j,k}^{(4)}\,\frac{\dot{x}^i\left(t\right)}{c}\,\frac{\dot{x}^j\left(t\right)}{c}\,\frac{\dot{x}^k\left(t\right)}{c}
- h_{0i,0}^{(4)} + {\cal O}\left(c^{-5}\right),  
\label{Geodesic_Equation3}
\end{eqnarray}
\end{widetext}

\noindent
where we have taken into account that in general $h_{0i}^{(2)}=h_{00}^{(3)}=h_{ij}^{(3)}=0$ \cite{MTW,Brumberg1991,IAU_Resolution1,Book_Clifford_Will,DSX1,DSX2}.  
The last term in (\ref{Geodesic_Equation3}), i.e. the term $h_{0i,0}^{(4)}$, is a peculiarity in the sense that this term is seemingly 
of the order ${\cal O}\left(c^{-5}\right)$,  
but by inspection of (\ref{Metric_4}) one realizes that the first integration (\ref{Integral_1}) of this term results into  
$4\,m_A\,\ve{a}_A/c^2$ which is of the order ${\cal O}\left(c^{-4}\right)$, and, therefore, cannot be neglected.  
Furthermore, the following relations have been used,  
\begin{eqnarray}
h_{00}^{(2)} &=& h^{00}_{(2)}\,,\; h_{ij}^{(2)} = h^{ij}_{(2)}\,,
\nonumber\\
h_{0i}^{(3)} &=& - h^{0i}_{(3)}\,,\; h_{0i}^{(4)} = - h^{0i}_{(4)}\,,
\nonumber\\
h_{00}^{(4)} &=& h^{00}_{(4)} - h^{00}_{(2)}\,h^{00}_{(2)}\,,\;
h_{ij}^{(4)} = h^{ij}_{(4)} + h^{ik}_{(2)}\,h^{kj}_{(2)}\,, 
\end{eqnarray}

\noindent
which result from $g_{\alpha\mu}\,g^{\mu \beta} = \delta_{\alpha}^{\beta} = {\rm diag}\left(+1,+1,+1,+1\right)$.  

The metric perturbations in (\ref{post_Newtonian_metric_B}) are functions of the field-points $(t,\ve{x})$,
while in the geodesic equation (\ref{Geodesic_Equation3}) the metric perturbations are of relevance at the coordinates of the photon $\ve{x}\left(t\right)$.
Consequently, the derivatives in (\ref{Geodesic_Equation3}) are taken along the light ray:
\begin{eqnarray}
h_{\alpha \beta, \mu}^{(n)} &=& \frac{\partial h_{\alpha \beta}^{(n)}\left(t,\ve{x}\right)}{\partial x^{\mu}}
\Bigg|_{\ve{x}=\ve{x}\mbox{\normalsize $\left(t\right)$}}\,,\quad n=2,3,4\,.
\label{geodesic_equation_3}
\end{eqnarray}

\noindent
The geodesic equation in 2PN approximation in (\ref{Geodesic_Equation3}) can be solved by iteration and allows to determine the 
coordinate velocity (first integration) and the light trajectory (second integration) up to terms of the order
${\cal O}\left(c^{-5}\right)$:
\begin{eqnarray}
\dot{\ve{x}}\left(t\right) &=& c\,\ve{\sigma} + \Delta \dot{\ve{x}}_{\rm 1PN} + \Delta \dot{\ve{x}}_{\rm 1.5PN} + \Delta \dot{\ve{x}}_{\rm 2PN}\,,  
\label{Light_Trajectory_2PN_A}
\\
\nonumber\\
\ve{x}\left(t\right) &=& \ve{x}_0 + c \left(t - t_0\right) \ve{\sigma} + \Delta \ve{x}_{\rm 1PN} + \Delta \ve{x}_{\rm 1.5PN}  
 +  \Delta \ve{x}_{\rm 2PN}\,.   
\nonumber\\
\label{Light_Trajectory_2PN_B}
\end{eqnarray}

\noindent
As mentioned above, the 1PN and 1.5PN terms in (\ref{Light_Trajectory_2PN_A}) and (\ref{Light_Trajectory_2PN_B}) have recently been determined 
in \cite{Zschocke_1PN} and \cite{Zschocke_15PN}, respectively, for the case of $N$ bodies in slow but otherwise arbitrary motion and the bodies may have  
arbitrary shape and inner structure and can be in arbitrary rotational motion. The aim of this investigation is to determine  
the 2PN terms for the case of one pointlike body in arbitrary slow-motion.  

%%%%%%%%%%%%%%%%%%%%%%%%%%%%%%%%%%%
\section{Metric in 2PN approximation for one body}\label{Section3}
%%%%%%%%%%%%%%%%%%%%%%%%%%%%%%%%%%%

%%%%%%%%%%%%%%%%%%%%%%%%%%%%%%%%%%%%%%%%%%%%%%%%%%%%%%%%%%%%%%%%%%%%%
%\begin{figure}[!ht]
\begin{figure}
\begin{center}
\includegraphics[scale=0.115]{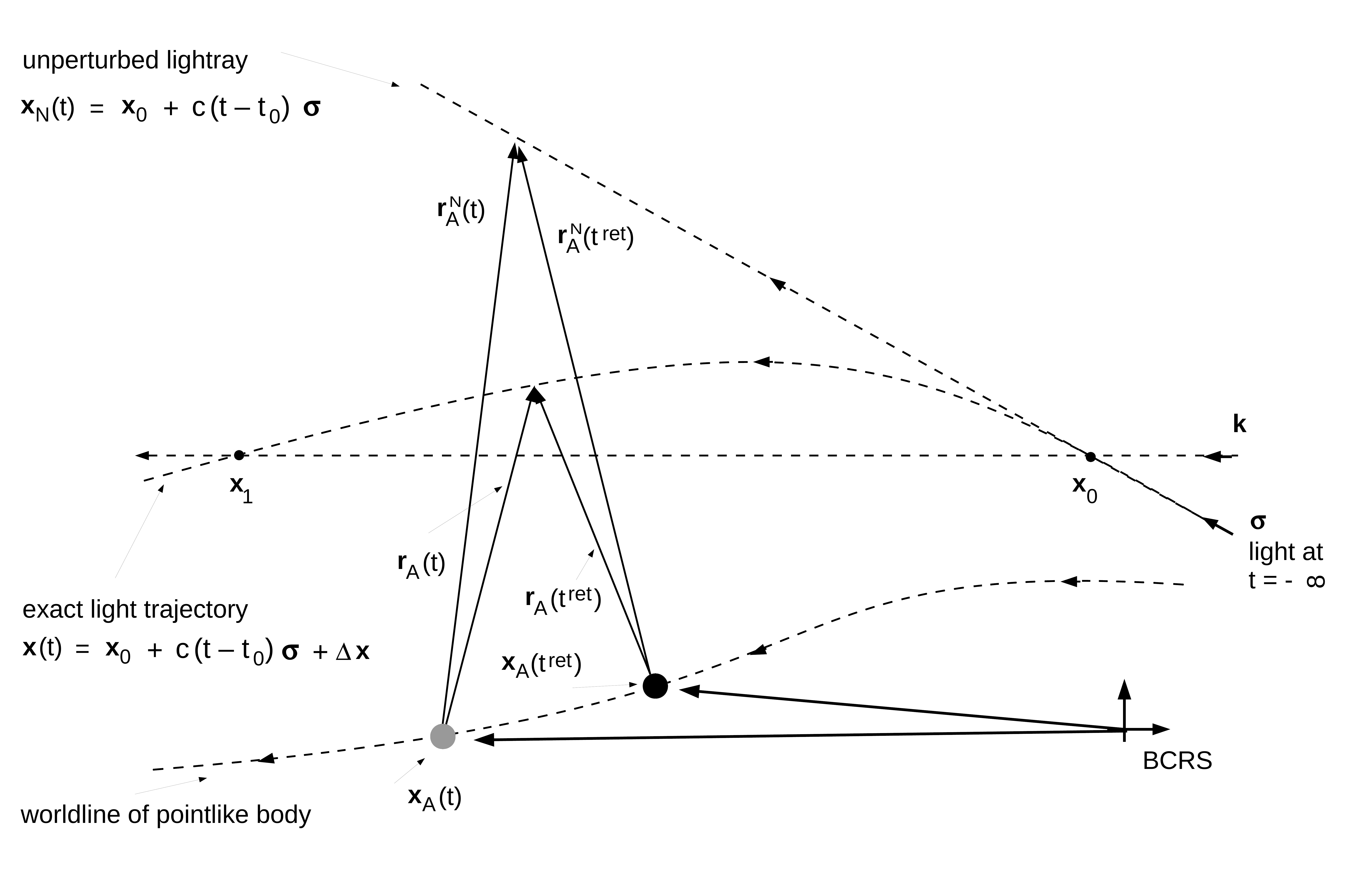}
\end{center}
\caption{A geometrical representation of the light trajectory $\ve{x}\left(t\right)$ of Eq.~(\ref{Second_Integration}) through the gravitational field  
of one pointlike massive body $A$ moving along an arbitrary worldline in slow motion $v_A \ll c$. At the same instant of coordinate time 
the body's position is $\ve{x}_A\left(t\right)$ (gray sphere). However, since gravitational action travels with the finite speed of light, 
the light ray at $\ve{x}\left(t\right)$ is influenced by the gravitational field generated by the body at its retarded position $\ve{x}_A\left(t^{\rm ret}\right)$ 
(black sphere). The spatial vector $\ve{r}_A\left(t^{\rm ret}\right)$ is defined by Eq.~(\ref{retarded_time_2}) and points from the massive body $A$  
at its retarded position toward the exact photon's position at instant $t$.}  
\label{Diagram}
\end{figure}
%%%%%%%%%%%%%%%%%%%%%%%%%%%%%%%%%%%%%%%%%%%%%%%%%%%%%%%%%%%%%%%%%%%%%

We shall assume that the one-body system is isolated (Fock-Sommerfeld boundary conditions), that means flatness of the metric at spatial  
infinity and the constraint of no-incoming gravitational radiation is imposed at Minkowskian past null infinity ${\cal J}_M^{-}$, which in 
terms of trace-reversed metric perturbation $\overline{h}^{\mu\nu} = \eta^{\mu\nu} - \sqrt{-g}\,g^{\mu\nu}$ read as follows  
\cite{Book_Fock,KlionerKopeikin1992,IAU_Resolution1,Kopeikin_Efroimsky_Kaplan,Radiation_Condition,Zschocke_Multipole_Expansion},  
\begin{eqnarray}
&& \hspace{-0.5cm} \lim_{r \rightarrow \infty \atop t  + \frac{r}{c} = {\rm const}}\,\overline{h}^{\mu \nu}\left(t,\ve{x}\right) \!=\! 0\,, 
\label{Asymptotic_1}
\\
\nonumber\\
&& \hspace{-0.5cm} \lim_{r \rightarrow \infty \atop t + \frac{r}{c} = {\rm const}}
\left(\frac{\partial}{\partial r} r\,\overline{h}^{\mu \nu}\left(t,\ve{x}\right) 
+ \frac{\partial}{\partial ct} \,r\,\overline{h}^{\mu \nu}\left(t,\ve{x}\right)\right) = 0\,,  
\label{Asymptotic_2}
\end{eqnarray}

\noindent
where $r = \left|\ve{x}\right|$.  
In addition, $r\,\partial_{\alpha}\,\overline{h}^{\mu \nu}$ should be bounded in this limit \cite{Book_Fock,Radiation_Condition}, 
that means any component of the metric tensor obeys the constraint  
\begin{eqnarray}
&& \hspace{-0.5cm} \lim_{r \rightarrow \infty \atop t  + \frac{r}{c} = {\rm const}}\,
\left|\frac{\partial \overline{h}^{\mu \nu}\left(t,\ve{x}\right)}{\partial x^{\alpha}}\right| < \frac{K}{r}\,,  
\label{Asymptotic_3}
\end{eqnarray}

\noindent
where $K>0$ is some positive number related to the total rest mass of the gravitational system. According to (\ref{post_Newtonian_metric_B})  
and Eqs.~(\ref{g00}) - (\ref{metric_w3}), the metric perturbations for the gravitational fields of one pointlike body in slow motion read:  
\begin{eqnarray}
h^{(2)}_{00}\left(t,\ve{x}\right) &=& + \frac{2\,m_A}{r_A\left(t\right)}\,,
\label{Metric_1}
\\
\nonumber\\
h^{(2)}_{ij}\left(t,\ve{x}\right) &=& + \frac{2\,m_A}{r_A\left(t\right)}\,\delta_{ij}\,,
\label{Metric_2}
\\
\nonumber\\
h^{(3)}_{0i}\left(t,\ve{x}\right) &=& - \frac{4\,m_A}{r_A\left(t\right)}\,\frac{v_A^i\left(t\right)}{c}\,,
\label{Metric_3}
\\
\nonumber\\
h^{(4)}_{0i}\left(t,\ve{x}\right) &=& + 4\,m_A\,\frac{a_A^i\left(t\right)}{c^2}\,,
\label{Metric_4}
\end{eqnarray}

\begin{widetext} 
\begin{eqnarray}
h^{(4)}_{00}\left(t,\ve{x}\right) &=& + \frac{4\,m_A}{r_A\left(t\right)}\,\frac{v^2_A\left(t\right)}{c^2} 
- \frac{m_A}{r_A\left(t\right)}\,\frac{\left(\ve{n}_A\left(t\right) \cdot \ve{v}_A\left(t\right)\right)^2}{c^2} 
- m_A\,\frac{\left(\ve{n}_A\left(t\right) \cdot \ve{a}_A\left(t\right)\right)}{c^2} 
- \frac{2\,m^2_A}{r^2_A\left(t\right)}\,,
\label{Metric_5}
\\
\nonumber\\
\nonumber\\
h^{(4)}_{ij}\left(t,\ve{x}\right) &=&
- \frac{m_A}{r_A\left(t\right)}\,\frac{\left(\ve{n}_A\left(t\right) \cdot \ve{v}_A\left(t\right)\right)^2}{c^2}\,\delta_{ij}
+ \frac{4\,m_A}{r_A\left(t\right)}\,\frac{v_A^i\left(t\right)}{c}\,\frac{v_A^j\left(t\right)}{c} 
- m_A\,\frac{\left(\ve{n}_A\left(t\right) \cdot \ve{a}_A\left(t\right)\right)}{c^2}\,\delta_{ij}
\nonumber\\
\nonumber\\
&& + \frac{m^2_A}{r^2_A\left(t\right)}\,\delta_{ij}
+ \frac{m^2_A}{r^2_A\left(t\right)}\,n_A^i\left(t\right)\,n_A^j\left(t\right),
\label{Metric_6}
\end{eqnarray}
\end{widetext} 

\noindent
where $m_A = G\,M_A/c^2$ and  
\begin{eqnarray}
\ve{r}_A\left(t\right) &=& \ve{x} - \ve{x}_A\left(t\right),
\label{vector_A}
\end{eqnarray}

\noindent
while its absolute value $r_A\left(t\right) = \left|\ve{r}_A\left(t\right)\right|$,  
and we introduce the unit-vector $\ve{n}_A\left(t\right) = \ve{r}_A\left(t\right)/r_A\left(t\right)$. 
The constraints (\ref{Asymptotic_1}) - (\ref{Asymptotic_3}) restrict the time-dependence of the acceleration so that  
it vanishes at past null infinity: $\lim_{t \rightarrow - \infty} \ve{a}_A\left(t\right) = 0$.  

The BCRS metric coefficients $h^{(2)}_{00}, h^{(2)}_{ij}, h^{(3)}_{0i}, h^{(4)}_{00}$ in the mass-monopole approximation for $N$ slowly moving bodies  
were given by Eqs.~(8) and (51) - (55) in \cite{IAU_Resolution1}. The same metric coefficients were also given by Eqs.~(39.63a) - (39.63c) in \cite{MTW}. In the  
limit of one monopole in slow-motion they agree with our metric coefficients in Eqs.~(\ref{Metric_1}), (\ref{Metric_2}), (\ref{Metric_3}) and (\ref{Metric_5}).  
We also notice that in the limit of one body $A$ at rest, the metric in (\ref{Metric_1}) - (\ref{Metric_6}) agrees 
with the metric in Eqs.~(25) in \cite{Klioner_Zschocke} if the body is assumed to be located at the origin of the global reference system.  
For further details consult \cite{Blanchet_Faye_Ponsot,Deng_Xie,Xie} and the Appendix \ref{Appendix_Metric}.  

The three-vector $\ve{x}$ in (\ref{vector_A}) is an arbitrary spatial  
field-point. But according to (\ref{geodesic_equation_3}), as soon as the partial derivatives in the geodesic equation (\ref{Geodesic_Equation3})  
are performed, the field-point $\ve{x}$ has to be identified with the exact spatial position of the light signal $\ve{x}\left(t\right)$, that means 
after all partial derivatives are performed we have  
\begin{eqnarray}
\ve{r}_A\left(t\right) &=& \ve{x}\left(t\right) - \ve{x}_A\left(t\right). 
\label{vector_B}
\end{eqnarray}

\noindent 
Of course, one has strictly to distinguish between (\ref{vector_A}) and (\ref{vector_B}), but nevertheless   
the same notation $\ve{r}_A\left(t\right)$ for these expressions is in use and will certainly not cause any confusion.  

%%%%%%%%%%%%%%%%%%%%%%%%%%%%%%%%%%%
\section{First integration of geodesic equation in 2PN approximation}\label{Section4}
%%%%%%%%%%%%%%%%%%%%%%%%%%%%%%%%%%%

The first integration of geodesic equation yields the coordinate velocity of the photon,
\begin{eqnarray}
\frac{\dot{\ve{x}}\left(t\right)}{c} &=& \int\limits_{- \infty}^{t} d c t \,\frac{\ddot{\ve{x}}\left(t\right)}{c^2}\,, 
\label{Integral_1} 
\end{eqnarray}

\noindent
where $\ddot{\ve{x}}\left(t\right)$ is given by (\ref{Geodesic_Equation3}) and the boundary condition (\ref{Initial_Boundary_Condition_2}) must be imposed.  
The geodesic equation (\ref{Geodesic_Equation3}) is solved by iteration, that means in first iteration the integration is performed along the unperturbed 
light ray and in the second iteration the integration proceeds along the light ray in 1PN approximation.  
Owing to the fact that the metric, thence the geodesic equation, depends on the arbitrary worldline of the body,   
a solution of geodesic equation is obtained by means of integration by parts with respect to coordinate time.  
One may show that, after a finite set of partial integrations, the remaining non-integrated terms of such an approach are terms beyond 
2PN approximation. The solution for the coordinate velocity of the photon is, first of all, given in terms of the  
spatial position of the massive body at coordinate time, $\ve{x}_A\left(t\right)$. Since gravitational action propagates with the finite speed of light, 
it is meaningful to reexpress this solution in terms of retarded time of the massive body's position $\ve{x}_A\left(t^{\rm ret}\right)$.  
The retarded time is defined by an implicit relation,
\begin{eqnarray}
t^{\rm ret} &=& t - \frac{r_A\left(t^{\rm ret}\right)}{c}\,,
\label{retarded_time}
\end{eqnarray}

\noindent
where $r_A\left(t^{\rm ret}\right) = \left|\ve{r}_A\left(t^{\rm ret}\right)\right|$ with
\begin{eqnarray}
\ve{r}_A\left(t^{\rm ret}\right) &=& \ve{x}\left(t\right) - \ve{x}_A\left(t^{\rm ret}\right),
\label{retarded_time_2}
\end{eqnarray}

\noindent
and $\ve{x}\left(t\right)$ being the exact photon trajectory. Further details are given in the Appendix \ref{Appendix_Retarded_Time}. 
 
Accordingly, we may define an impact vector of the incident light ray associated with the body's position at retarded instant of time, given by
\begin{eqnarray}
\ve{d}_A\left(t^{\rm ret}\right) &=& \ve{\sigma} \times \left(\ve{r}_A\left(t^{\rm ret}\right) \times \ve{\sigma}\right),
\label{retarded_time_3}
\end{eqnarray}

\noindent
and weak gravitational field means
\begin{eqnarray}
m_A &\ll& d_A\left(t^{\rm ret}\right)\,,
\label{retarded_time_4}
\end{eqnarray}

\noindent
where $d_A\left(t^{\rm ret}\right) = \left|\ve{d}_A\left(t^{\rm ret}\right)\right|$. For grazing light rays the impact vector at the retarded position
equals the radius of the massive body, while in general it will be larger: $d_A\left(t^{\rm ret}\right) \ge P_A$.
It should also be remarked that the solution for the coordinate velocity of the photon takes the most simple form in terms of $\ve{x}_A\left(t^{\rm ret}\right)$.  
In this way one obtains the following 2PN solution for the photon's coordinate velocity in the field of one arbitrarily moving pointlike body:  
\begin{widetext} 
\begin{eqnarray}
\frac{\dot{\ve{x}}\left(t\right)}{c} &=& \ve{\sigma}
+ m_A\,\ve{A}_1\left(\ve{r}_A^{\rm 1PN}\left(t^{\rm ret}\right)\right)
+ m_A\,\ve{A}_2\left(\ve{r}_A^{\rm N}\left(t^{\rm ret}\right),\ve{v}_A\left(t^{\rm ret}\right)\right)
+ m_A^2\,\ve{A}_3\left(\ve{r}_A^{\rm N}\left(t^{\rm ret}\right)\right) + {\cal O}\left(c^{-5}\right)\,,
\label{First_Integration} 
\\
\nonumber\\ 
\nonumber\\ 
\ve{A}_1\left(\ve{x}\right) &=&
- 2\,\left(\frac{\ve{\sigma} \times \left(\ve{x} \times \ve{\sigma}\right)}
{x \left(x - \ve{\sigma} \cdot \ve{x}\right)}
+ \frac{\ve{\sigma}}{x} \right),  
\label{Vectorial_Function_A1}
\\
\nonumber\\
\nonumber\\
\ve{A}_2\left(\ve{x},\ve{v}\right) &=& 2\,\frac{\ve{\sigma} \times \left(\ve{x} \times \ve{\sigma}\right)}
{x \left(x - \ve{\sigma} \cdot \ve{x}\right)}\,
\frac{\ve{\sigma} \cdot \ve{v}}{c}
+ \frac{4}{x}\,\frac{\ve{v}}{c}
+ 2\,\frac{\ve{\sigma} \times \left(\ve{x} \times \ve{\sigma}\right)}
{x^2}\,\frac{\ve{\sigma} \cdot \ve{v}}{c}
- 2\,\frac{\ve{\sigma}}{x^2}\,
\frac{\ve{x} \cdot \ve{v}}{c}
- 2\,\frac{\ve{\sigma} \times \left(\ve{x} \times \ve{\sigma}\right)}
{x^2\,\left(x - \ve{\sigma} \cdot \ve{x}\right)}\,  
\frac{\left(\ve{\sigma} \times \left(\ve{x} \times \ve{\sigma}\right) \right) \cdot \ve{v}}{c} + \ve{\epsilon}_1\,,
\nonumber\\
\label{Vectorial_Function_A2}
\\
\nonumber\\
\ve{A}_3\left(\ve{x}\right) &=& - \frac{1}{2}\,\frac{\ve{\sigma} \cdot \ve{x}}{x^4}\,\ve{x}
+ 8\,\frac{\ve{\sigma} \times \left(\ve{x} \times \ve{\sigma}\right)}{x^2\,\left(x - \ve{\sigma} \cdot\ve{x} \right)}\,
+ 4\,\frac{\ve{\sigma} \times \left(\ve{x} \times \ve{\sigma}\right)}{x\,\left(x - \ve{\sigma} \cdot \ve{x} \right)^2}\,
- 4\,\frac{\ve{\sigma}}{x\,\left(x - \ve{\sigma} \cdot \ve{x} \right)}
+ \frac{9}{2}\,\frac{\ve{\sigma}}{x^2}
- \frac{15}{4}\,\left(\ve{\sigma} \cdot \ve{x}\right)\,
\frac{\ve{\sigma} \times \left(\ve{x} \times \ve{\sigma}\right)}{x^2\,\left|\ve{\sigma} \times \ve{x}\right|^2}
\nonumber\\
\nonumber\\
&& - \frac{15}{4}\,\frac{\ve{\sigma} \times \left(\ve{x} \times \ve{\sigma}\right)}{\left|\ve{\sigma} \times \ve{x}\right|^3}
\left(\arctan \frac{\ve{\sigma} \cdot \ve{x}}{\left|\ve{\sigma} \times \ve{x}\right|} + \frac{\pi}{2}\right), 
\label{Vectorial_Function_A3}
\end{eqnarray}
\end{widetext} 

\noindent
where the arguments of the vectorial functions are given in Appendix \ref{Appendix_1PN_Solution}. 
One may demonstrate that in case of body at rest (\ref{First_Integration}) agrees with \cite{Brumberg1987,Brumberg1991,KlionerKopeikin1992,Klioner_Zschocke},  
and the terms up to order ${\cal O}\left(c^{-4}\right)$ agree with \cite{KopeikinSchaefer1999,Zschocke_15PN}. In view of the complexity of the 2PN solution,  
all those terms in (\ref{Vectorial_Function_A2}) have been combined in some small parameter $\ve{\epsilon}_1$ given by Eq.~(\ref{epsilon_1}) 
and which has been estimated by Eq.~(\ref{estimate_epsilon_1}):  
\begin{eqnarray}  
\left|\ve{\epsilon}_1\left(\ve{r}_A^{\rm N}\left(t^{\rm ret}\right),\ve{v}_A\left(t^{\rm ret}\right)\right)\right| 
&\le& \frac{10}{d^{\rm N}_A\left(t^{\rm ret}\right)} \,\frac{v_A^2\left(t^{\rm ret}\right)}{c^2}\,,
\label{estimate_epsilon1}
\end{eqnarray}

\noindent
which amounts to be less than $\;\;m_A \left|\ve{\epsilon}_1\right| < 0.1\,{\rm nas}\;\;$ for grazing light rays at Jupiter and even less for 
all the other Solar System bodies.   
We emphasize that the solution in (\ref{First_Integration}) does not depend on the acceleration of the massive body, as long as it   
is given in terms of the retarded position of the massive body. This fact  
is related to the case of an arbitrarily moving and radiating electron in classical electrodynamics where the Li\'enard-Wiechert potentials 
do not depend on the  
acceleration of the electron if the worldline of the electron is expressed in terms of its retarded position \cite{Jackson_Electrodynamics,Feynman}.  

Finally, we notice that the vectorial function $\ve{A}_1$ in Eq.~(\ref{Vectorial_Function_A1}) agrees with Eq.~(46) in \cite{Klioner_Zschocke} 
and the vectorial function $\ve{A}_3$ in Eq.~(\ref{Vectorial_Function_A3}) agrees with Eq.~(48) in \cite{Klioner_Zschocke} (in general theory of relativity $\alpha=\beta=\gamma=\epsilon=1$).  

%%%%%%%%%%%%%%%%%%%%%%%%%%%%%%%%%%%
\section{Second integration of geodesic equation in 2PN approximation}\label{Section5}
%%%%%%%%%%%%%%%%%%%%%%%%%%%%%%%%%%%

The second integration of geodesic equation yields the light trajectory of the light signal,
\begin{eqnarray}
\ve{x}\left(t\right) &=& \int\limits_{t_0}^{t} d c t \,\frac{\dot{\ve{x}}\left(t\right)}{c}\,,
\label{Integral_2}
\end{eqnarray}

\noindent
where $\dot{\ve{x}}\left(t\right)$ is given by (\ref{First_Integration}) and the boundary condition (\ref{Initial_Boundary_Condition_1}) must be imposed. 
Note that (\ref{First_Integration}) is given in terms of retarded time, but in order to proceed with the integration in (\ref{Integral_2}) all terms  
in (\ref{First_Integration}) must have to be reexpressed in terms of coordinate time by means of relations (\ref{retarded_time_B}) - (\ref{retarded_time_D}).  
Like in case of the first integration in (\ref{Integral_1}), the second integration in (\ref{Integral_2}) is performed by iteration.  
Furthermore, since the worldline of the body remains arbitrary, $\ve{x}_A\left(t\right)$, the integration is performed by means of integration by parts.  
In this way, one obtains the light trajectory in terms of the spatial position of the massive body at coordinate time, $\ve{x}_A\left(t\right)$, but   
can be rewritten in terms of retarded time of the position of massive body, $\ve{x}_A\left(t^{\rm ret}\right)$, which is also from the physical point of view 
more appropriate because gravitational action travels with the finite speed of light.  
Altogether, we obtain the following 2PN solution for the photon's trajectory in the  
field of one arbitrarily moving pointlike body, for an illustration see Fig.~\ref{Diagram}:  
\begin{widetext} 
\begin{eqnarray}
\ve{x}\left(t\right) &=& \ve{x}_0 + c \left(t - t_0\right) \ve{\sigma}
+ \,m_A\,\bigg(\ve{B}_1\left(\ve{r}_A^{\rm 1PN}\left(t^{\rm ret}\right)\right)
- \ve{B}_1\left(\ve{r}_A^{\rm 1PN}\left(t_0^{\rm ret}\right)\right) \bigg)
\nonumber\\
\nonumber\\
&& \hspace{-1.5cm} + \,m_A\,\bigg(\ve{B}_2\left(\ve{r}_A^{\rm N}\left(t^{\rm ret}\right),\ve{v}_A\left(t^{\rm ret}\right)\right)
- \ve{B}_2\left(\ve{r}_A^{\rm N}\left(t_0^{\rm ret}\right),\ve{v}_A\left(t_0^{\rm ret}\right)\right) \bigg)
+ \,m_A^2\,
\bigg(\ve{B}_3\left(\ve{r}_A^{\rm N}\left(t^{\rm ret}\right)\right) - \ve{B}_3\left(\ve{r}_A^{\rm N}\left(t_0^{\rm ret}\right)\right) \bigg) 
+ {\cal O}\left(c^{-5}\right),  
\nonumber\\ 
\label{Second_Integration}
\\
\nonumber\\ 
\ve{B}_1\left(\ve{x}\right) &=& -\,2\, \frac{\ve{\sigma} \times \left(\ve{x} \times \ve{\sigma}\right)}{x - \ve{\sigma} \cdot \ve{x}}
+\,2\,\ve{\sigma}\,\ln \left(x - \ve{\sigma} \cdot \ve{x}\right), 
\label{Vectorial_Function_B1}
\\
\nonumber\\
\ve{B}_2\left(\ve{x},\ve{v}\right) &=& 
2\,\frac{\ve{\sigma} \times \left(\ve{x} \times \ve{\sigma}\right)}{x - \ve{\sigma} \cdot \ve{x}}\,\frac{\ve{\sigma} \cdot \ve{v}}{c}
-\,2\,\frac{\ve{v}}{c}\,\ln \left(x - \ve{\sigma} \cdot \ve{x}\right) +\,2\,\frac{\ve{v}}{c} + \ve{\epsilon}_2\,,
\label{Vectorial_Function_B2}
\\
\nonumber\\
\ve{B}_3\left(\ve{x}\right) &=& 4\,\frac{\ve{\sigma}}{x - \ve{\sigma} \cdot \ve{x}}
+\,4\,\frac{\ve{\sigma} \times \left(\ve{x} \times \ve{\sigma}\right)}{\left(x - \ve{\sigma} \cdot \ve{x}\right)^2}
+\,\frac{1}{4}\,\frac{\ve{x}}{x^2}
-\,\frac{15}{4}\,\frac{\ve{\sigma}}{\left|\ve{\sigma} \times \ve{x}\right|} \,
\arctan \frac{\ve{\sigma} \cdot \ve{x}}{\left|\ve{\sigma} \times \ve{x}\right|}  
\nonumber\\
\nonumber\\
&& -\,\frac{15}{4}\,\left(\ve{\sigma} \cdot \ve{x}\right) \frac{\ve{\sigma} \times \left(\ve{x} \times \ve{\sigma}\right)}
{\left|\ve{\sigma} \times \ve{x}\right|^3} \left(\arctan \frac{\ve{\sigma} \cdot \ve{x}}{\left|\ve{\sigma} \times \ve{x}\right|} + \frac{\pi}{2}\right),  
\label{Vectorial_Function_B3}
\end{eqnarray}
\end{widetext}

\noindent
where the arguments of the vectorial functions are given in Appendix \ref{Appendix_1PN_Solution}.  
One may demonstrate that in case of body at rest (\ref{Second_Integration}) agrees with \cite{Brumberg1987,Brumberg1991,KlionerKopeikin1992,Klioner_Zschocke},
and the terms up to order ${\cal O}\left(c^{-4}\right)$ agree with \cite{KopeikinSchaefer1999,Zschocke_15PN}. In view of the complexity of the 2PN solution,  
all those terms in (\ref{Vectorial_Function_B2}) have been combined in some small parameter $\ve{\epsilon}_2$ given by Eq.~(\ref{epsilon_2})
and which has been estimated by Eq.~(\ref{estimate_epsilon_2}):
\begin{eqnarray}
\left|\ve{\epsilon}_2\left(\ve{r}_A^{\rm N}\left(t^{\rm ret}\right),\ve{v}_A\left(t^{\rm ret}\right)\right)\right| &\le& \frac{v_A^2\left(t^{\rm ret}\right)}{c^2}\,
\nonumber\\ 
\nonumber\\ 
&& \hspace{-4.5cm} \times \sqrt{\frac{4\,\left(r^{\rm N}_A\left(t^{\rm ret}\right)\right)^2}{\left(d^{\rm N}_A\left(t^{\rm ret}\right)\right)^2}
+ \ln^2 \left(r^{\rm N}_A\left(t^{\rm ret}\right) - \ve{\sigma} \cdot \ve{r}^{\rm N}_A\left(t^{\rm ret}\right)\right)} \,,  
\label{estimate_epsilon2}
\end{eqnarray}

\noindent
which for grazing light rays at Jupiter ($d^{\rm N}_A = 7.15 \times 10^7\,{\rm m}$) and an observer nearby the Earth 
($r^{\rm N}_A = 0.59 \times 10^{12}\,{\rm m}$) amounts to be $m_A\,\left|\ve{\epsilon}_2\right| < 10^{-4}\,{\rm m}$ and even less 
for all the other Solar system bodies.  

We emphasize that the solution in (\ref{Second_Integration}) does not depend on the acceleration of the massive body, as long as the 
solution is given in terms of the retarded position of the massive body.  
This important fact resembles the case of an arbitrarily moving electron, where the Li\'enard-Wiechert potential does not depend
on the acceleration as long as the worldline of the electron is given in terms of its retarded position \cite{Jackson_Electrodynamics,Feynman}.

We notice that the vectorial function $\ve{B}_1$ in Eq.~(\ref{Vectorial_Function_B1}) agrees with Eq.~(50) in \cite{Klioner_Zschocke}
and the vectorial function $\ve{B}_3$ in Eq.~(\ref{Vectorial_Function_B3}) agrees with Eq.~(51) in \cite{Klioner_Zschocke} (in general theory of relativity $\alpha=\beta=\gamma=\epsilon=1$).

\section{Observable effects}\label{Observables}  

In this section we briefly consider the observable effects of total light deflection and time delay which are of upmost relevance for astrometry and 
belong to the classical tests of relativity.  

\subsection{Total light deflection}

The total light deflection of a light signal propagating through the gravitational field of one arbitrarily moving body 
is defined by the angle between the coordinate light velocity at $t \rightarrow \pm \infty$. From (\ref{First_Integration}) we first of all obtain    
up to terms of the order ${\cal O}\left(c^{-5}\right)$:  
\begin{eqnarray}
\lim_{t \rightarrow - \infty}\,\frac{\dot{\ve{x}}\left(t\right)}{c} &\equiv& \ve{\sigma}\,, 
\label{total_light_deflection_sigma} 
\\ 
\nonumber\\
\nonumber\\
\lim_{t \rightarrow + \infty}\,\frac{\dot{\ve{x}}\left(t\right)}{c} &\equiv& \ve{\nu}  
\nonumber\\
&& \hspace{-2.0cm} = \ve{\sigma} - 4\,m_A\,\frac{\ve{d}_A\left(t^{\rm ret}\right)}{\left(d_A\left(t^{\rm ret}\right)\right)^2} 
\left(1 - \frac{\ve{\sigma} \cdot \ve{v}_A\left(t^{\rm ret}\right)}{c}\right)  
\nonumber\\
\nonumber\\
&& \hspace{-2.0cm} - 8\,m_A^2 \frac{\ve{\sigma}}{\left(d^{\rm N}_A\left(t^{\rm ret}\right)\right)^2}  
- \frac{15}{4}\,\pi\,m_A^2\,\frac{\ve{d}^{\rm N}_A\left(t^{\rm ret}\right)}{\left(d_A^{\rm N}\left(t^{\rm ret}\right)\right)^3}  
\nonumber\\
\nonumber\\
&& \hspace{-2.0cm} + 8\,m_A^2\,\frac{\ve{d}^{\rm N}_A\left(t^{\rm ret}\right)}{\left(d^{\rm N}_A\left(t^{\rm ret}\right)\right)^4} 
\left(r_A^{\rm N}\left(t^{\rm ret}\right) + \ve{\sigma} \cdot \ve{r}_A^{\rm N}\left(t^{\rm ret}\right)\right). 
\label{total_light_deflection_nu} 
\end{eqnarray} 

\noindent  
In the limit of one body at rest and located at the origin of reference system, the expression in (\ref{total_light_deflection_nu}) agrees 
with Eq.~(64) in \cite{Klioner_Zschocke}.   
The impact vector $\ve{d}_A\left(t^{\rm ret}\right)$ and the impact vector $\ve{d}^{\rm N}_A\left(t^{\rm ret}\right)$ are related to each other subject  
to (\ref{Solution_1PN_A}). Then, from (\ref{total_light_deflection_nu}) one obtains for the total light deflection  
up to terms of the order ${\cal O}\left(c^{-5}\right)$:  
\begin{widetext} 
\begin{eqnarray}
\left|\ve{\sigma} \times \ve{\nu}\right| &=& \lim_{t \rightarrow + \infty} \frac{4\,m_A}{d^{\rm N}_A\left(t^{\rm ret}\right)}
\bigg[1 - \frac{\ve{\sigma}\cdot\ve{v}_A\left(t^{\rm ret}\right)}{c}
- 2\,m_A\,\frac{r_A\left(t^{\rm ret}_0\right) 
+ \ve{\sigma}\cdot \ve{r}_A\left(t^{\rm ret}_0\right)}{\left(d^{\rm N}_A\left(t^{\rm ret}\right)\right)^2}
\frac{\ve{d}^{\rm N}_A\left(t^{\rm ret}_0\right) \cdot \ve{d}^{\rm N}_A\left(t^{\rm ret}\right)}{\left(d^{\rm N}_A\left(t^{\rm ret}_0\right)\right)^2}
+ \frac{15}{16}\,\pi\,\frac{m_A}{d_A^{\rm N}\left(t^{\rm ret}\right)}\bigg],
\nonumber\\ 
\label{Light_Deflection}
\end{eqnarray}
\end{widetext} 

\noindent
where $t_0^{\rm ret} = t_0 - r_A\left(t_0^{\rm ret}\right)/c$ and
\begin{eqnarray}
\ve{r}_A\left(t_0^{\rm ret}\right) &=& \ve{x}\left(t_0\right) - \ve{x}_A\left(t_0^{\rm ret}\right),
\label{retarded_time_1}
\end{eqnarray}

\noindent
and terms $\left|\ve{\sigma} \times \ve{\epsilon}_1\right|$ have been omitted in (\ref{Light_Deflection}) in view of the estimate in (\ref{estimate_epsilon1}).
Furthermore, in (\ref{Light_Deflection}) the impact vector at $t^{\rm ret}_0$ and $t^{\rm ret}$ of the unperturbed light ray has been used:
\begin{eqnarray}
\ve{d}^{\rm N}_A\left(t^{\rm ret}_0\right) &=& \ve{\sigma} \times \bigg(\left(\ve{x}_0 - \ve{x}_A\left(t^{\rm ret}_0\right)\right) \times \ve{\sigma} \bigg),
\label{Impact_Vector_1}
\\
\nonumber\\
\ve{d}^{\rm N}_A\left(t^{\rm ret}\right) &=& \ve{\sigma} \times \bigg(\left(\ve{x}_0 - \ve{x}_A\left(t^{\rm ret}\right)\right) \times \ve{\sigma} \bigg),
\label{Impact_Vector_2}
\end{eqnarray}

\noindent
which in the case of a motionless body at the origin of coordinate system coincides with the impact vector defined by Eq.~(55) in \cite{Klioner_Zschocke}.
The expression in (\ref{Light_Deflection}) depends on the direction of the light ray $\ve{\sigma}$, on the coordinates of the light source
$\ve{x}_0, t_0$ and on the mass, position and velocity of the massive body $m_A, \ve{x}_A, \ve{v}_A$ and it
generalizes the corresponding 2PN expression for a body at rest \cite{Brumberg1987,Brumberg1991,Klioner_Zschocke}, 
cf. Eq.~(3.2.44) in \cite{Brumberg1991} or Eq.~(65) in \cite{Klioner_Zschocke}.
The occurrence of the third term in the brackets in (\ref{Light_Deflection}) is caused by the fact that the
total light deflection, which is a coordinate-independent observable, is expressed in terms of coordinate-dependent quantities.
This assertion can be shown by introducing a coordinate independent impact vector similar to the one given by Eq.~(57) in \cite{Klioner_Zschocke}.
But, as emphasized above, the use of concrete reference systems is inevitable in real astrometric data reduction.

\subsection{Time-delay}

A light signal which propagates through the curved space of a massive body takes a longer time to travel from one space-time point to 
another space-time point compared to the flat Minkowskian space.  
Let's assume the light source and the observer to be located at $\left(\ve{x}_0, t_0\right)$ and $\left(\ve{x}_1, t_1\right)$, respectively,  
and to be at rest with respect to the global reference system, and we may define a spatial distance $R = \left|\ve{x}_1 - \ve{x}_0\right|$.  
Then, from (\ref{Second_Integration}) one obtains the following expression for the time delay up to terms of the order ${\cal O}\left(c^{-5}\right)$:    
\begin{widetext} 
\begin{eqnarray}
c\,\left(t_1 - t_0\right) &=& R - 2\,m_A\,
\left(\frac{\ve{\sigma} \cdot \ve{v}_A\left(t_1^{\rm ret}\right)}{c} - \frac{\ve{\sigma} \cdot \ve{v}_A\left(t_0^{\rm ret}\right)}{c} \right) 
- 2\,m_A\,\left(1 - \frac{\ve{\sigma}\cdot\ve{v}_A\left(t_1^{\rm ret}\right)}{c}\right) 
\ln \left(r_A\left(t_1^{\rm ret}\right) - \ve{\sigma} \cdot \ve{r}_A\left(t_1^{\rm ret}\right) + 2\,m_A\right)
\nonumber\\
\nonumber\\
&& \hspace{5.0cm} + 2\,m_A\,\left(1 - \frac{\ve{\sigma}\cdot\ve{v}_A\left(t_0^{\rm ret}\right)}{c}\right)
\ln \left(r_A\left(t_0^{\rm ret}\right) - \ve{\sigma} \cdot \ve{r}_A\left(t_0^{\rm ret}\right) + 2\,m_A\right),   
\label{Shapiro}
\end{eqnarray}
\end{widetext} 
 
\noindent
where $t_1^{\rm ret} = t_1 - r_A\left(t_1^{\rm ret}\right)/c$; the terms $\ve{\sigma} \cdot \ve{\epsilon}_2$ were neglected 
in view of the estimate in (\ref{estimate_epsilon2}). The expression in (\ref{Shapiro})  
generalizes the corresponding 2PN expression for one monopole at rest \cite{Brumberg1987,Brumberg1991,Klioner_Zschocke,Moyer} and  
it generalizes the expression in Eqs.~(146) - (148) in \cite{Zschocke_15PN} which is valid for arbitrarily moving monopoles but in 1.5PN approximation.

%%%%%%%%%%%%%%%%%%%%%%%%%%%%%%%%%%%
\section{Summary and Outlook}\label{Section6}
%%%%%%%%%%%%%%%%%%%%%%%%%%%%%%%%%%%

Present-day astrometry has reached a level of a few micro-arcseconds in angular determination of celestial objects and prospective astrometry aims at  
sub-micro-arcsecond or even nano-arcsecond level of precision. Associated therewith is the precise determination of light trajectories through the warped 
space-time of the Solar System as one central issue in relativistic astrometry. An exact solution for the light ray is, however, not possible 
because of the involved structure of the metric of the Solar System and one has, therefore, to resort on approximation schemes.  
The gravitational fields of the Solar System are weak, $m_A/P_A \ll 1$, and the velocities of the bodies are slow, $v_A/c \ll 1$, so that an expansion  
of the metric tensor of the Solar System in inverse powers of the speed of light becomes meaningful as given by Eq.~(\ref{post_Newtonian_metric_B}), 
\begin{eqnarray}
g_{\alpha \beta} &=& \eta_{\alpha \beta} + h^{(2)}_{\alpha\beta} + h^{(3)}_{\alpha\beta} + h^{(4)}_{\alpha\beta}\,,  
\label{Summary_1}
\end{eqnarray}

\noindent 
up to terms of the order ${\cal O} \left(c^{-5}\right)$. 
This so-called post-Newtonian expansion (weak-field slow-motion approximation)  
implicitly assumes that all retardations are small, which is well-justified inside the near-zone of the Solar System.  
A corresponding expansion of the light trajectory is given by Eq.~(\ref{Introduction_6}) and reads 
\begin{eqnarray}
\ve{x}\left(t\right) &=& \ve{x}_0 + c \left(t - t_0\right) \ve{\sigma} + \Delta \ve{x}_{\rm 1PN} + \Delta \ve{x}_{\rm 1.5PN}
+ \Delta \ve{x}_{\rm 2PN}\,,
\nonumber\\
\label{Summary_2}
\end{eqnarray}
  
\noindent
up to terms of the order ${\cal O} \left(c^{-5}\right)$. 
One of the most intricate problems in the relativistic theory of light propagation concerns the impact of the motion of the massive bodies on light trajectory. In  
recent investigations \cite{Zschocke_1PN,Zschocke_15PN} the 1PN and 1.5PN terms, $\Delta \ve{x}_{\rm 1PN}$ and $\Delta \ve{x}_{\rm 1.5PN}$, have been determined 
for the case of $N$ arbitrarily moving bodies having full mass-multipole and spin-multipole structure. The rapid advance in astrometric measurements  
enforces one to account for post-post-Newtonian effects $\Delta \ve{x}_{\rm 2PN}$ in the theory of light propagation as well. The 2PN terms in  
(\ref{Summary_2}) are only known for the case of one monopole at rest, first been determined in \cite{Brumberg1987,Brumberg1991} and later been confirmed  
within several ongoing investigations \cite{KlionerKopeikin1992,Klioner_Zschocke,Zschocke_15PN,Deng_Xie,Teyssandier,Hees_Bertone_Poncin_Lafitte_2014b}.  
But little is known about these terms in (\ref{Summary_2}) for the case of moving bodies. So far, the only investigation in 2PN approximation regarding  
light trajectory in the field of moving bodies has been performed in \cite{Bruegmann2005} which was, however, not intended for light propagation inside the 
Solar System.  

In our investigation, the problem of light propagation in the field of one arbitrarily moving pointlike monopole has been considered. 
Especially, an analytical solution in  
post-post-Newtonian approximation for coordinate velocity $\dot{\ve{x}}\left(t\right)$ and trajectory $\ve{x}\left(t\right)$ of the light ray is presented.  
According to the recommendations of IAU \cite{IAU_Resolution1} the metric is given in terms of harmonic coordinates. Because of the fact that the  
worldline $\ve{x}_A\left(t\right)$ of the massive body is arbitrarily, an integration of the geodesic equation in (\ref{Geodesic_Equation3}) is  
only possible by means of integration by parts. The first integration (\ref{Integral_1}) and the second integration (\ref{Integral_2}) has been  
performed in terms of coordinate time. In this respect one has to keep in mind that the post-Newtonian expansion of the metric (\ref{Summary_1}) 
and of the light ray (\ref{Summary_2}) inherits that all retardations are small, but they are not negligible. Instead, the fact remains that gravitational 
action travels with the speed of light also inside the near-zone of the Solar system. The phrase {\it smallness of retardation effects} in the Solar System 
means that a series  
expansion of the retarded time is meaningful, as given by Eqs.~(\ref{retarded_time_A2}) - (\ref{retarded_time_D}).  
By means of these relations the solution, first of all given in terms of the instantaneous position of the body $\ve{x}_A\left(t\right)$, can be expressed 
in terms of the retarded position of the body $\ve{x}_A\left(t^{\rm ret}\right)$, where the first integration and the second integration of geodesic equation 
adopt the most simple form, as given by Eqs.~(\ref{First_Integration}) and (\ref{Second_Integration}), respectively.   
The expressions for the observables of total light deflection and of Shapiro time delay are given by Eqs.~(\ref{Light_Deflection}) and (\ref{Shapiro}).  

The case of $N$ arbitrarily moving pointlike bodies is rather involved and needs special consideration. But in view of the fact that the impact of  
two-body effects on light deflection is less than $0.1\,{\rm nas}$ in the Solar System \cite{Deng_Xie}, one might assert that the 2PN light trajectory  
in the field of $N$ pointlike monopoles in arbitrary slow-motion can be obtained from our solution just by a summation over $N$ individual bodies,  
at least for an envisaged accuracy on nas-level. These aspects will have to be scrutinized within more detailed prospective analyses.

%%%%%%%%%%%%%%%%%%%%%%%%%%%%%%%%%%%
\section{Acknowledgment}
%%%%%%%%%%%%%%%%%%%%%%%%%%%%%%%%%%%

This work was supported by the Deutsche Forschungsgemeinschaft (DFG).

\appendix

\section{Notation}\label{Appendix_Notation}  

Throughout the article the following notation is in use. 

\begin{itemize}
\item $G$ is the Newtonian constant of gravitation.
\item $c$ is the vacuum speed of light.
\item $M_A$ denotes the rest mass of body $A$. 
\item $m_A = G\,M_A/c^2$ is the Schwarzschild radius.  
\item $P_A$ denotes the equatorial radius of body $A$.  
\item $v_A$ denotes the orbital velocity of massive body $A$.   
\item $\displaystyle 1\,\muas = \frac{\pi}{180 \times 60 \times 60}\,10^{-6}\,{\rm rad} \simeq 4.85 \times 10^{-12}\,{\rm rad}$.\\ 
\item $\displaystyle 1\,{\rm nas} = \frac{\pi}{180 \times 60 \times 60}\,10^{-9}\,{\rm rad} \simeq 4.85 \times 10^{-15}\,{\rm rad}$.
\item Lower case Latin indices take values 1,2,3.
\item Lower case Greek indices take values 0,1,2,3.
\item The three-dimensional coordinate quantities (three-vectors) referred to
the spatial axes of the reference system are in boldface: $\ve{a}$.
\item The contravariant components of three-vectors $a^{i} = \left(a^1,a^2,a^3\right)$.
\item The contravariant components of four-vectors $a^{\mu} = \left(a^0,a^1,a^2,a^3\right)$.
\item The absolute value of a three-vector.
$a = |\ve{a}| = \sqrt{a^1\,a^1+a^2\,a^2+a^3\,a^3}$.
\item The scalar product of two three-vectors.
$\ve{a}\,\cdot\,\ve{b}=\delta_{ij}\,a^i\,b^j=a^i\,b^i$ with Kronecker delta $\delta_{ij}$.
\item The vector product of two three-vectors.
reads $\left(\ve{a}\times\ve{b}\right)^i=\varepsilon_{ijk}\,a^j\,b^k$  
with Levi-Civita symbol $\varepsilon_{ijk}$.
\end{itemize}

\section{Retarded time}\label{Appendix_Retarded_Time}  

Gravitational action travels with the finite speed of light which implies that the gravitational field at some field point $\ve{x}$ is 
generated by the pointlike body at its position $\ve{x}_A\left(t^{\rm ret}\right)$ at the retarded instant of time defined by  
\begin{eqnarray}
t^{\rm ret} &=& t - \frac{r_A\left(t^{\rm ret}\right)}{c}\,,  
\label{retarded_time_A1}
\end{eqnarray}

\noindent
where $r_A\left(t^{\rm ret}\right) = \left|\ve{x} - \ve{x}_A\left(t^{\rm ret}\right)\right|$. 
In the near-zone of the Solar System \cite{MTW,Poisson_Will,Kopeikin_Efroimsky_Kaplan} one may assume that all retardations are small,  
hence a series expansion of (\ref{retarded_time_A1}) becomes meaningful, 
\begin{eqnarray}
t^{\rm ret} &=& t - \frac{r_A\left(t\right)}{c} - \frac{\ve{r}_A\left(t\right) \cdot \ve{v}_A\left(t\right)}{c^2} + {\cal O}\left(c^{-3}\right)\,,  
\label{retarded_time_A2}
\end{eqnarray}

\noindent  
which will later be used for the series expansion of the metric tensor.
Using (\ref{retarded_time_A2}) and the series expansion of the retarded position of the body which up to terms of the order ${\cal O}\left(c^{-3}\right)$ reads 
\begin{eqnarray}
\ve{x}_A\left(t^{\rm ret}\right) &=& \ve{x}_A\left(t\right) + \dot{\ve{x}}_A\left(t\right) \frac{\left(t^{\rm ret} - t\right)}{1!}  
+ \ddot{\ve{x}}_A\left(t\right) \frac{\left(t^{\rm ret} - t\right)^2}{2!}\,, 
\nonumber\\
\label{retarded_position}
\end{eqnarray}
 
\noindent
we find the following relations: 
\begin{widetext} 
\begin{eqnarray}
\ve{r}_A\left(t^{\rm ret}\right) &=& \ve{r}_A\left(t\right) + \frac{\ve{v}_A\left(t\right)}{c}\,r_A\left(t\right)
+ \frac{\ve{v}_A\left(t\right)}{c}\,\frac{\ve{r}_A\left(t\right) \cdot \ve{v}_A\left(t\right)}{c}
- \frac{1}{2}\,\frac{\ve{a}_A\left(t\right)}{c}\,\frac{r^2_A\left(t\right)}{c} + {\cal O}\left(c^{-3}\right),
\label{retarded_time_B}
\\
\nonumber\\
\nonumber\\
r_A\left(t^{\rm ret}\right) &=& r_A\left(t\right)
\bigg(1 + \frac{\ve{r}_A\left(t\right) \cdot \ve{v}_A\left(t\right)}{c\,r_A\left(t\right)} + \frac{1}{2}\,\frac{v_A^2\left(t\right)}{c^2}
+ \frac{1}{2}\,\frac{\left(\ve{v}_A\left(t\right) \cdot \ve{r}_A\left(t\right)\right)^2}{c^2\,r_A^2\left(t\right)}
- \frac{1}{2}\,\frac{\ve{r}_A\left(t\right) \cdot \ve{a}_A\left(t\right)}{c^2}\bigg) + {\cal O}\left(c^{-3}\right),
\label{retarded_time_C}
\\
\nonumber\\
\nonumber\\
\frac{\ve{v}_A\left(t^{\rm ret}\right)}{c} &=& \frac{\ve{v}_A\left(t\right)}{c} - \frac{\ve{a}_A\left(t\right)}{c}\,\frac{r_A\left(t\right)}{c} 
+ {\cal O}\left(c^{-3}\right), 
\label{retarded_time_D}
\end{eqnarray}
\end{widetext} 

\noindent
where $\ve{v}_A\left(t\right) = \dot{\ve{x}}_A\left(t\right)$ and $\ve{a}_A\left(t\right) = \ddot{\ve{x}}_A\left(t\right)$ is the velocity and acceleration of  
the body, respectively. These relations agree with Eqs.~(47) - (49) in \cite{Zschocke_Soffel} up to the term proportional to the acceleration $\ve{a}_A$.  
These relations have been obtain for $\ve{r}_A\left(t^{\rm ret}\right)=\ve{x}-\ve{x}_A\left(t^{\rm ret}\right)$, but we notice that 
the relations in (\ref{retarded_time_A2}) and (\ref{retarded_time_B}) - (\ref{retarded_time_D}) remain its validity for 
$\ve{r}_A\left(t^{\rm ret}\right)=\ve{x}\left(t\right)-\ve{x}_A\left(t^{\rm ret}\right)$, because they root on the expansion in (\ref{retarded_position}). 

The series-expansions in (\ref{retarded_time_B}) - (\ref{retarded_time_D}) are useful as long as the retardations are small
which is well-justified in the near-zone of the Solar System. It especially constraints the accelerations,
\begin{eqnarray}
\frac{a_A\left(t\right)\,r_A\left(t\right)}{c^2} &\ll& \frac{v_A\left(t\right)}{c} \ll 1\;,
\label{constraint_acceleration}
\end{eqnarray}

\noindent
for any moment of time.

\section{2PN metric for one arbitrarily moving body}\label{Appendix_Metric} 

For our intention we need the 2PN metric in harmonic gauge (\ref{Harmonic_Gauge}) for the case of one arbitrarily but slowly moving pointlike monopole.
The 2PN metric contains terms proportional to $G$ and terms proportional to $G^2$, which are considered in what follows.

\subsection{Metric coefficients proportional to $G$}\label{Appendix_Metric_G1}

The terms proportional to $G$ can easily be obtained from
the metric for one arbitrarily moving pointlike monopole in post-Minkowskian approximation, which has been given by Eq.~(10) in \cite{KopeikinSchaefer1999},
by Eq.~(11) in \cite{KopeikinMashhoon2002}, and also by Eq.~(43) in \cite{Zschocke_Soffel}:
\begin{widetext} 
\begin{eqnarray}
h_{\mu \nu}^{\left(M\right)} \left(t^{\rm ret},\ve{x}\right) &=&  
\frac{4\,m_A}{\gamma_A\left(t^{\rm ret}\right)\,\left(r_A\left(t^{\rm ret}\right) -
\frac{\displaystyle \ve{v}_A\left(t^{\rm ret}\right) \cdot \ve{r}_A\left(t^{\rm ret}\right)}{\displaystyle c}\right)}\,
\left(\frac{u^A_{\mu}\left(t^{\rm ret}\right)}{c}\,\frac{u^A_{\nu}\left(t^{\rm ret}\right)}{c}+\frac{\eta_{\mu \nu}}{2}\right),
\label{Arbitrarily_Moving_Body_4}
\end{eqnarray}
\end{widetext} 

\noindent
where $\gamma^{-1}_A\left(t^{\rm ret}\right)=\sqrt{1 - v^2_A\left(t^{\rm ret}\right)/c^2}$ is the Lorentz factor and $\left(M\right)$ denotes monopole.
The vector pointing from the retarded position $\ve{x}_A\left(t^{\rm ret}\right)$ of the body $A$ towards the field-point $\ve{x}$ reads:
\begin{eqnarray}
\ve{r}_A\left(t^{\rm ret}\right) &=& \ve{x} - \ve{x}_A\left(t^{\rm ret}\right). 
\label{Retarded_Position_5}
\end{eqnarray}

\noindent
Let us note that in (\ref{Arbitrarily_Moving_Body_4}) we present the covariant components of the metric perturbations,
while in \cite{KopeikinSchaefer1999,KopeikinMashhoon2002,Zschocke_Soffel} the contravariant components have been used.
Accordingly, the covariant components of the four-velocity of the body are
$u^A_{\mu}\left(t^{\rm ret}\right) = \gamma\left(t^{\rm ret}\right)\left(- c, \ve{v}_A\left(t^{\rm ret}\right)\right)$,
and $\ve{v}_A\left(t^{\rm ret}\right)$ is the three-velocity of the body in the global system.
Let us also draw the attention to the fact, that the metric tensor in (\ref{Arbitrarily_Moving_Body_4}) does not depend on the acceleration but
only on the velocity of body $A$ because it is given in terms of retarded time, see also comment below Eq.~(\ref{estimate_epsilon1}).

The metric in (\ref{Arbitrarily_Moving_Body_4}) is valid for an arbitrarily moving body which could  
even be in ultra-relativistic motion. We are interested in the case of a slowly-moving body, and a corresponding  
series expansion of (\ref{Arbitrarily_Moving_Body_4}) in terms of the small parameter $v_A/c \ll 1$ yields up to order ${\cal O}\left(c^{-5}\right)$:  
\begin{widetext}
\begin{eqnarray}
h_{00}^{\left(M\right)} \left(t^{\rm ret},\ve{x}\right) &=& + \frac{2\,m_A}{r_A\left(t^{\rm ret}\right)}
\left(1 + \frac{\ve{v}_A\left(t^{\rm ret}\right)\cdot\ve{r}_A\left(t^{\rm ret}\right)}{c\,r_A\left(t^{\rm ret}\right)}
+ \frac{\left(\ve{v}_A\left(t^{\rm ret}\right)\cdot\ve{r}_A\left(t^{\rm ret}\right)\right)^2}
{c^2 r_A^2\left(t^{\rm ret}\right)}
+ \frac{3}{2} \frac{v_A^2\left(t^{\rm ret}\right)}{c^2}\right), 
\label{Comparison_5}
\\
\nonumber\\
\nonumber\\
h_{0i}^{\left(M\right)} \left(t^{\rm ret},\ve{x}\right) &=& - \frac{4\,m_A}{r_A\left(t^{\rm ret}\right)}
\frac{v_A^i\left(t^{\rm ret}\right)}{c}\,
\left(1 + \frac{\ve{v}_A\left(t^{\rm ret}\right)\cdot\ve{r}_A\left(t^{\rm ret}\right)}{c\,r_A\left(t^{\rm ret}\right)} \right), 
\label{Comparison_10}
\\
\nonumber\\
\nonumber\\
h_{ij}^{\left(M\right)} \left(t^{\rm ret},\ve{x}\right) &=&
+ \frac{2\,m_A}{r_A\left(t^{\rm ret}\right)}\,\delta_{ij}
\left(1 + \frac{\ve{v}_A\left(t^{\rm ret}\right)\cdot\ve{r}_A\left(t^{\rm ret}\right)}{c\,r_A\left(t^{\rm ret}\right)}
+ \frac{\left(\ve{v}_A\left(t^{\rm ret}\right)\cdot\ve{r}_A\left(t^{\rm ret}\right)\right)^2}
{c^2\,r_A^2\left(t^{\rm ret}\right)}
- \frac{1}{2}\,\frac{v_A^2\left(t^{\rm ret}\right)}{c^2} \right)
\nonumber\\
\nonumber\\
&& + \frac{4\,m_A}{r_A\left(t^{\rm ret}\right)}\,
\frac{v_A^i\left(t^{\rm ret}\right)\,v_A^j\left(t^{\rm ret}\right)}{c^2}\,. 
\label{Comparison_15}
\end{eqnarray}
\end{widetext}

\noindent
The retarded time-argument in (\ref{Comparison_5}) - (\ref{Comparison_15}) has to be replaced by the global coordinate time using the relations 
in (\ref{retarded_time_B}) - (\ref{retarded_time_D}).  

Before going further, one has to realize that the acceleration of some body $A$ is proportional to $G$ according to the equations of motion  
for $N$ pointlike bodies,  
\begin{eqnarray}
\ve{a}_A \left(t\right) &=& - G \sum\limits_{B \neq A}^{N-1} M_B\,\frac{\ve{r}_A\left(t\right) - \ve{r}_B\left(t\right)}{r_{AB}^3}
+ {\cal O}\left(c^{-2}\right)\!,  
\label{equations_of_motion}
\end{eqnarray}

\noindent
where the terms of order ${\cal O}\left(c^{-2}\right)$ are given by the Einstein-Infeld-Hoffmann equations 
\cite{Einstein_Infeld_Hoffmann,MTW,Brumberg1991,Book_Clifford_Will,Kopeikin_Efroimsky_Kaplan}.  
Here, however, we cannot use the equations of motion (\ref{equations_of_motion})  
because we consider the metric of one body $A$ in arbitrary motion and the physical origin of the motion of the body
is not relevant for the moment being. Especially, we do not have some kind of equations of motion like in an $N$-body system (just imagine accelerating rockets
tied to that body). Therefore, we are enforced to keep the acceleration terms explicitly in Eqs.~(\ref{retarded_time_B}) - (\ref{retarded_time_D}).
Of course, if one would go back and consider an $N$-body system under the influence of their mutual gravitational interaction, then one could make use
of the equations of motion (\ref{equations_of_motion}) and then such an acceleration term would appear as term of the order $G^2$ in the metric tensor.
According to these considerations, to order $G$ we obtain:
\begin{eqnarray}
h^{(2)}_{00}\left(t,\ve{x}\right) &=& + \frac{2\,m_A}{r_A\left(t\right)}\,,
\label{Appendix_Metric_1}
\\
\nonumber\\
h^{(2)}_{ij}\left(t,\ve{x}\right) &=& + \frac{2\,m_A}{r_A\left(t\right)}\,\delta_{ij}\,,
\label{Appendix_Metric_2}
\\
\nonumber\\
h^{(3)}_{0i}\left(t,\ve{x}\right) &=& - \frac{4\,m_A}{r_A\left(t\right)}\,\frac{v_A^i\left(t\right)}{c}\,,
\label{Appendix_Metric_3}
\end{eqnarray}

\noindent
while $h_{0i}^{(2)}=h_{00}^{(3)}=h_{ij}^{(3)} = 0$ and

\begin{widetext} 
\begin{eqnarray}
h^{(4)\,G}_{00}\left(t,\ve{x}\right) &=& + \frac{4\,m_A}{r_A\left(t\right)}\,\frac{v^2_A\left(t\right)}{c^2}
- \frac{m_A}{r_A\left(t\right)}\,\frac{\left(\ve{n}_A\left(t\right) \cdot \ve{v}_A\left(t\right)\right)^2}{c^2}
- \frac{m_A}{r_A\left(t\right)}\,\frac{\left(\ve{r}_A\left(t\right) \cdot \ve{a}_A\left(t\right)\right)}{c^2}\,,
\label{Appendix_Metric_4}
\\
\nonumber\\
h^{(4)\,G}_{0i}\left(t,\ve{x}\right) &=& + 4\,m_A\,\frac{a_A^i\left(t\right)}{c^2}\,,
\label{Appendix_Metric_5}
\\
\nonumber\\
h^{(4)\,G}_{ij}\left(t,\ve{x}\right) &=&
- \frac{m_A}{r_A\left(t\right)}\,\frac{\left(\ve{n}_A\left(t\right) \cdot \ve{v}_A\left(t\right)\right)^2}{c^2}\,\delta_{ij}
+ \frac{4\,m_A}{r_A\left(t\right)}\,\frac{v_A^i\left(t\right)}{c}\,\frac{v_A^j\left(t\right)}{c}
- \frac{m_A}{r_A\left(t\right)}\,\frac{\left(\ve{r}_A\left(t\right) \cdot \ve{a}_A\left(t\right)\right)}{c^2}\,\delta_{ij}\,.
\label{Appendix_Metric_6}
\end{eqnarray}
\end{widetext} 

\noindent
We recognize that in
(\ref{Appendix_Metric_4}) - (\ref{Appendix_Metric_6}) there are terms proportional to the acceleration of the body.
The metric in Eqs.~(\ref{Appendix_Metric_1}) - (\ref{Appendix_Metric_6}) agrees with the metric given by Eqs.~(7.2a) - (7.2c) in \cite{Blanchet_Faye_Ponsot}
for all terms proportional to $G$ and up to the order ${\cal O}\left(c^{-5}\right)$. But we notice that in Eqs.~(7.2a) - (7.2c) in \cite{Blanchet_Faye_Ponsot}
there are no acceleration terms, because they have been rewritten by means of the equations of motion of an $N$-body system (\ref{equations_of_motion}),
cf. Eq.~(3.11) in \cite{Blanchet_Faye_Ponsot} and the text above that equation. 

Another point to mention concerns the expression in (\ref{Appendix_Metric_5}). For an $N$-body system it is a strict law that there are no terms to power $c^{-4}$
in $g_{0i}$ \cite{Brumberg1991,MTW,Poisson_Will,Kopeikin_Efroimsky_Kaplan}, because in an $N$-body system, instead of (\ref{Appendix_Metric_5}), we would have
a summation over all bodies,
\begin{eqnarray}
h^{(4)\,G}_{0i}\left(t,\ve{x}\right) &=& + \frac{4\,G}{c^4} \sum\limits_{A=1}^N M_A\,a_A^i\left(t\right)
\nonumber\\
\nonumber\\
&=& + \frac{4\,G}{c^4} \frac{d}{d t} P^i\left(t\right) = {\cal O}\left(c^{-6}\right),
\end{eqnarray}

\noindent
where $P^i\left(t\right)=\sum\limits_{A=1}^N M_A\,v_A^i\left(t\right)$ is the total Newtonian momentum of the $N$-body system, which is strictly conserved
to order ${\cal O}\left(c^{-2}\right)$, that means $\displaystyle \frac{d}{dt}\,P^i\left(t\right) = {\cal O}\left(c^{-2}\right)$ \cite{Poisson_Will}.
Therefore, in an $N$-body system there is in fact no term to power $c^{-4}$ in $g_{0i}$ \cite{Brumberg1991,MTW,Poisson_Will,Kopeikin_Efroimsky_Kaplan}. But
in our case of one single body which moves along an arbitrary worldline without to resort on the equations of motion (\ref{equations_of_motion}), there is no
conservation of total Newtonian momentum, hence we have to keep that term in (\ref{Appendix_Metric_5}). Nevertheless, there is an important difference
regarding the acceleration terms: in 2PN approximation the acceleration term in (\ref{Appendix_Metric_5}) would disappear for an $N$-body system, while
the acceleration term in (\ref{Appendix_Metric_4}) and (\ref{Appendix_Metric_6}) could be rewritten in a form proportional to $G^2$, but they remain to be of the
order ${\cal O}\left(c^{-4}\right)$, hence they would not disappear for an $N$-body system in 2PN approximation.

\subsection{Metric coefficients proportional to $G^2$}\label{Appendix_Metric_G2}

The metric of a system of two pointlike bodies under the influence of their mutual gravitational interaction has been determined in 2.5PN approximation in  
\cite{Blanchet_Faye_Ponsot}, that means $g_{00}$, $g_{i0}$ and $g_{ij}$ up to terms of the order ${\cal O}\left(c^{-8}\right)$, ${\cal O}\left(c^{-7}\right)$, and 
${\cal O}\left(c^{-6}\right)$, respectively. Recently, the metric of $N$ pointlike bodies has been determined in \cite{Deng_Xie} in 2PN approximation for the
light rays, that is: $g_{\alpha\beta}$ up to terms of the order ${\cal O}\left(c^{-5}\right)$. In order to find all terms proportional to $G^2$, we may   
issue the results from Ref.~\cite{Deng_Xie}, but have to take the limit $M_B \rightarrow 0$ for all bodies
except body $A$. In this way we obtain from Eqs.~(47) - (49) (with $\alpha=\beta=\gamma=1$) in \cite{Deng_Xie}:
\begin{eqnarray}
h^{(4)\,G^2}_{00}\!\left(t,\ve{x}\right) &=&
- \frac{2\,m_A^2}{r^2_A\left(t\right)}\,,
\label{Appendix_Metric_G2_1}
\\
\nonumber\\
h^{(4)\,G^2}_{ij}\!\left(t,\ve{x}\right) &=&
\frac{m^2_A}{r^2_A\left(t\right)}\,\delta_{ij}
+ \frac{m^2_A}{r^2_A\left(t\right)}\,n_A^i\!\left(t\right) n_A^j\!\left(t\right).  
\label{Appendix_Metric_G2_2}
\end{eqnarray}

\noindent
The metric perturbations $h^{(4)}_{\alpha\beta}$ are given by  
\begin{eqnarray}
h^{(4)}_{\alpha\beta} &=& h^{(4)\,G}_{\alpha \beta} + h^{(4)\,G^2}_{\alpha \beta}\,,
\label{Metric_G_GG}
\end{eqnarray}

\noindent
with $h^{(4)\,G}_{\alpha \beta}$ given by (\ref{Appendix_Metric_4}) - (\ref{Appendix_Metric_6})   
and $h^{(4)\,G^2}_{\alpha \beta}$ given by (\ref{Appendix_Metric_G2_1}) and (\ref{Appendix_Metric_G2_2}).  

Let us draw the attention to the fact, that if we insert the equations of motion (\ref{equations_of_motion}) into the last term in (\ref{Appendix_Metric_4}), then
we would get the next-to-last term in Eq.~(47) in \cite{Deng_Xie}. Similarly, if we insert the equations of motion (\ref{equations_of_motion}) into the last term
in (\ref{Appendix_Metric_6}), then we would get the next-to-last term in the second line of Eq.~(49) in \cite{Deng_Xie}. But we emphasize again, that we are
not allowed to do such kind of replacements (which in \cite{Blanchet_Faye_Ponsot} were called order-reduced form of the metric, cf. text above Eq.~(3.11) ibid),
because we do not consider an $N$-body system but a system of one body which moves arbitrarily along its worldline without taking resort on the equations of motion
in (\ref{equations_of_motion}).

\subsection{Collection of all terms} 

Collecting all metric coefficients in Eqs.~(\ref{Appendix_Metric_1}) - (\ref{Appendix_Metric_6}) and  
in Eqs.~(\ref{Appendix_Metric_G2_1}) - (\ref{Metric_G_GG}), the post-post Newtonian metric for light rays 
can also expressed in terms of so-called potentials ($w$, $w_i$, $\tau_{ij}$) in the following form,  
cf. Eqs.~(A1) - (A3) in \cite{2PN_Lightray_Metric1} or Eq.~(2) in \cite{2PN_Lightray_Metric2}:  
\begin{widetext} 
 \begin{eqnarray}
g_{00}\left(t,\ve{x}\right) &=& - 1 + 2\,w\left(t,\ve{x}\right) - 2\,w^2\left(t,\ve{x}\right) + {\cal O}\left(c^{-5}\right),   
\label{g00} 
\\
\nonumber\\
g_{0i}\left(t,\ve{x}\right) &=& - 4\,w_i\left(t,\ve{x}\right) + {\cal O}\left(c^{-5}\right),  
\label{g0i} 
\\
\nonumber\\
g_{ij}\left(t,\ve{x}\right) &=& \left(1 + 2\,w\left(t,\ve{x}\right) + 2\,w^2\left(t,\ve{x}\right)\right) \delta_{ij} 
+ 4\,\tau_{ij}\left(t,\ve{x}\right) + {\cal O}\left(c^{-5}\right),  
\label{gij} 
\end{eqnarray}
\end{widetext} 

\noindent
where the potentials read   
\begin{widetext}
\begin{eqnarray}
w\left(t,\ve{x}\right) &=& \frac{m_A}{r_A\left(t\right)} + \frac{3}{2}\,\frac{m_A}{r_A\left(t\right)}\, \frac{v_A^2\left(t\right)}{c^2}
+ \frac{1}{2}\,\frac{m_A}{c^2}\,\frac{d^2}{dt^2}\,r_A\left(t\right) 
\nonumber\\
\nonumber\\
&=& \frac{m_A}{r_A\left(t\right)} + 2\,\frac{m_A}{r_A\left(t\right)}\,\frac{v_A^2\left(t\right)}{c^2}
- \frac{1}{2}\,\frac{m_A}{r_A\left(t\right)}\,\frac{\left(\ve{n}_A\left(t\right) \cdot \ve{v}_A\left(t\right)\right)^2}{c^2}
- \frac{1}{2}\,m_A\,\frac{\left(\ve{n}_A\left(t\right) \cdot \ve{a}_A\left(t\right)\right)}{c^2}\,,  
\label{metric_w1} 
\\
\nonumber\\
\nonumber\\
w_i\left(t,\ve{x}\right) &=& \frac{m_A}{r_A\left(t\right)}\,\frac{v_A^i\left(t\right)}{c} - m_A\,\frac{a_A^i\left(t\right)}{c^2}\,,
\label{metric_w2}
\\
\nonumber\\
\nonumber\\
\tau_{ij}\left(t,\ve{x}\right) &=& - \frac{m_A}{r_A\left(t\right)}\,\frac{v_A^2\left(t\right)}{c^2}\,\delta_{ij} 
+ \frac{m_A}{r_A\left(t\right)}\,\frac{v_A^i\left(t\right)}{c} \frac{v_A^j\left(t\right)}{c}
- \frac{1}{4}\,\frac{m_A^2}{r_A^2\left(t\right)}\,\delta_{ij} 
+ \frac{1}{4}\,\frac{m_A^2}{r_A^2\left(t\right)}\,n_A^i\left(t\right)\,n_A^j\left(t\right), 
\label{metric_w3}  
\end{eqnarray}
\end{widetext}

\noindent
recalling $r_A\left(t\right) = \left|\ve{x} - \ve{x}_A\left(t\right)\right|$.

\section{Light trajectory in 1PN approximation}\label{Appendix_1PN_Solution}

The Newtonian and the first post-Newtonian solution for the light ray appears as argument in the vectorial functions of
the 2PN solution in (\ref{First_Integration}) and (\ref{Second_Integration}). In this Appendix we will present these expressions.
In Newtonian approximation we have
\begin{eqnarray}
\ve{r}_A^{\rm N}\left(t^{\rm ret}\right) &=& \ve{x}_0 + c \left(t-t_0\right) \ve{\sigma} - \ve{x}_A\left(t^{\rm ret}\right).  
\label{Solution_N_A}
\end{eqnarray}

\noindent
Furthermore, the light trajectory in the field of one arbitrarily moving body in the first post-Newtonian approximation can be
obtained from \cite{Zschocke_15PN} by means of relations (\ref{retarded_time_B}) - (\ref{retarded_time_C}) and reads:
\begin{eqnarray}
\ve{r}_A^{\rm 1PN}\!\left(t^{\rm ret}\right) &=& \ve{r}_A^{\rm N}\left(t^{\rm ret}\right)  
+ 2\,m_A\,\ve{\sigma} \ln \frac{r_A^{\rm N}\left(t^{\rm ret}\right) - \ve{\sigma} \cdot \ve{r}_A^{\rm N}\left(t^{\rm ret}\right)}
{r_A^{\rm N}\left(t_0^{\rm ret}\right) - \ve{\sigma} \cdot \ve{r}_A^{\rm N}\left(t_0^{\rm ret}\right)} 
\nonumber\\
\nonumber\\
&& \hspace{-2.0cm} - 2\,m_A \! \left(\!\!\frac{\ve{\sigma} \times \left(\ve{r}_A^{\rm N}\left(t^{\rm ret}\right) \times \ve{\sigma}\right)}
{r_A^{\rm N}\!\left(t^{\rm ret}\right) - \ve{\sigma} \cdot \ve{r}_A^{\rm N}\left(t^{\rm ret}\right)}
- \frac{\ve{\sigma} \times \left(\ve{r}_A^{\rm N}\left(t_0^{\rm ret}\right) \times \ve{\sigma}\right)}
{r_A^{\rm N}\!\left(t_0^{\rm ret}\right) - \ve{\sigma} \cdot \ve{r}_A^{\rm N}\left(t_0^{\rm ret}\right)}\!\!\right). 
\nonumber\\ 
\label{Solution_1PN_A}
\end{eqnarray}

\section{The expressions $\ve{\epsilon}_1$ and $\ve{\epsilon}_2$}\label{Epsilon_Terms} 

The term $\ve{\epsilon}_1$ in Eq.~(\ref{Vectorial_Function_A2}) reads as follows: 
\begin{widetext}  
\begin{eqnarray} 
\ve{\epsilon}_1 \left(\ve{x},\ve{v}\right) &=& 
- \frac{v^2}{c^2}\,\frac{\ve{\sigma} \times \left(\ve{x} \times \ve{\sigma}\right)}{x - \ve{\sigma} \cdot \ve{x}}\,\frac{1}{x}  
- 2\,\left(\frac{\ve{v} \cdot \ve{x}}{c\,x}\right)^2\,
\frac{\ve{\sigma} \times \left(\ve{x} \times \ve{\sigma}\right)}{x - \ve{\sigma} \cdot \ve{x}}\,\frac{1}{x}  
- 2\,\left(\frac{\ve{\sigma} \cdot \ve{v}}{c}\right)^2\,\frac{\ve{\sigma} \times \left(\ve{x} \times \ve{\sigma}\right)}{x - \ve{\sigma} \cdot \ve{x}}\,\frac{1}{x} 
\nonumber\\
\nonumber\\
&& + 4\,\left(\frac{\ve{\sigma} \cdot \ve{v}}{c}\right) \, \left(\frac{\ve{v} \cdot \ve{x}}{c\,x}\right) \, 
\frac{\ve{\sigma} \times \left(\ve{x} \times \ve{\sigma}\right)}{x - \ve{\sigma} \cdot \ve{x}}\,\frac{1}{x}  
+ 4\,\frac{\ve{v}}{c}\,\left(\frac{\ve{v} \cdot \ve{x}}{c\,x}\right) \,\frac{1}{x} 
- 4\,\frac{\ve{v}}{c}\,\left(\frac{\ve{\sigma} \cdot \ve{v}}{c}\right)\,\frac{1}{x} 
\nonumber\\
\nonumber\\
&& - \frac{v^2}{c^2}\,\frac{\ve{\sigma}}{x} 
- 2\, \left(\frac{\ve{v} \cdot \ve{x}}{c\,x}\right)^2\,\frac{\ve{\sigma}}{x} 
+ 2\, \left(\frac{\ve{\sigma} \cdot \ve{v}}{c}\right)^2\,\frac{\ve{\sigma}}{x}\,.  
\label{epsilon_1} 
\end{eqnarray} 
\end{widetext}  

\noindent
By means of the angles   
\begin{eqnarray}
\left(\frac{\ve{v} \cdot \ve{x}}{c\,x}\right) &=& \frac{v}{c}\,\cos \alpha \,, 
\label{angle_alpha}
\\
\nonumber\\
\left(\frac{\ve{\sigma} \cdot \ve{v}}{c}\right) &=& \frac{v}{c}\,\cos \beta \,,  
\label{angle_beta}
\end{eqnarray} 

\noindent 
one obtains up to terms of the order 
$\displaystyle {\cal O}\left(\frac{m_A}{x}\frac{v^2}{c^2}\right)$ the following upper limit: 
\begin{eqnarray}
m_A\,\left|\ve{\epsilon}_1\left(\ve{x},\ve{v}\right) \right| &\le& 18\,\frac{m_A}{\left|\ve{\sigma} \times \ve{x}\right|}\,\frac{v^2}{c^2} 
+ 9\,\frac{m_A}{x}\,\frac{v^2}{c^2}\,,  
\label{estimate_epsilon_1}
\end{eqnarray}

\noindent 
where for the first and second term on the r.h.s. in (\ref{estimate_epsilon_1}) we have used 
\begin{eqnarray}
\left| - 1 - 2\,\cos^2 \alpha - 2\,\cos^2 \beta + 4\,\cos \alpha\;\cos \beta\right| &\le& 9\,,   
\end{eqnarray}

\noindent
and 
\begin{widetext}
\begin{eqnarray}
\sqrt{16\left(\cos \alpha - \cos \beta\right)^2 - 8 \,\cos \beta \left(\cos \alpha - \cos \beta\right) \left(1 + 2\,\cos^2 \alpha - 2\,\cos^2 \beta\right) 
+ \left(1 + 2\,\cos^2 \alpha - 2\,\cos^2 \beta\right)^2} &\le& 9\,, 
\end{eqnarray}
\end{widetext}

\noindent 
respectictively. The term $\ve{\epsilon}_2$ in Eq.~(\ref{Vectorial_Function_B2}) reads as follows:  
\begin{eqnarray}
\ve{\epsilon}_2 \left(\ve{x},\ve{v}\right) &=& 
- \frac{v^2}{c^2}\,\frac{\ve{\sigma} \times \left(\ve{x} \times \ve{\sigma}\right)}{x - \ve{\sigma} \cdot \ve{x}} 
+ \frac{v^2}{c^2}\, \ve{\sigma}\,\ln \left(x - \ve{\sigma} \cdot \ve{x}\right).  
\nonumber\\
\label{epsilon_2}
\end{eqnarray}

\noindent
The absolute value of $\ve{\epsilon}_2$ can be estimated by  
\begin{eqnarray}
\left|\ve{\epsilon}_2\left(\ve{x},\ve{v}\right) \right| &\le& \frac{v^2}{c^2}\, 
\sqrt{\frac{4\,x^2}{\left|\ve{\sigma} \times \ve{x}\right|^2} + \ln^2 \left(x - \ve{\sigma} \cdot \ve{x}\right)}\,. 
\label{estimate_epsilon_2}
\end{eqnarray}

\noindent  

%%%%%%%%%%%%%%%%%%%%%%%%%%%%%%%%%%%%%%%

\end{document}